\documentclass[preprint]{aastex}
\usepackage{amsmath}
\usepackage[colorlinks,citecolor=blue]{hyperref}
\usepackage{xcolor}
\usepackage{bm}
\def\coi    {$^{12}$CO}
\def\coii   {$^{13}$CO}
\def\coiii  {C$^{18}$O}
\def\kms    {km s$^{-1}$}
\def\Msun   {M$_{\odot}$}
\def\Hm     {H$_2$}
\def\HII    {H{\scriptsize~II}}
\def\Lsun   {L$_{\odot}$}

\def\nht    {NH$_3$}
\def\nhti   {NH$_3$(1,1)}
\def\nhtii  {NH$_3$(2,2)}

\def\Tex    {$T_{\mathrm{ex}}$}
\def\Trot   {$T_{\mathrm{rot}}$}
\def\Tkin   {$T_{\mathrm{kin}}$}
\def\Tmb    {$T_{\mathrm{mb}}$}
\def\eqTex  {T_{\mathrm{ex}}}
\def\eqTrot {T_{\mathrm{rot}}}
\def\eqTkin {T_{\mathrm{kin}}}
\def\eqTmb  {T_{\mathrm{mb}}}
\def\etamb  {$\eta_{\mathrm{mb}}$}
\def\vpeak  {$v_{\mathrm{peak}}$}
\def\eqbr   {\mathrm{b/r}}

\shorttitle{NH$_3$ and CO Outflows}
\shortauthors{Li et al.}

\begin{document}


\title{Ammonia and CO Outflow around 6.7 GHz Methanol Masers}
%Star Formation around 6.7 GHz Methanol Masers: Ammonia and CO Outflow Properties


%%[A4]
\author{F. C. Li\altaffilmark{1,2,3}, Y. Xu\altaffilmark{1,3}, Y. W. Wu\altaffilmark{4}, J. Yang\altaffilmark{1,3}, D. R. Lu\altaffilmark{1}, K. M. Menten\altaffilmark{5}, C. Henkel\altaffilmark{5,6}}
\email{xuye@pmo.ac.cn; lifc@pmo.ac.cn}


\altaffiltext{1}{Purple Mountain Observatory, Chinese Academy of Sciences, Nanjing 210008, China}
\altaffiltext{2}{University of Chinese Academy of Sciences, Beijing 100049, China}
\altaffiltext{3}{Key Laboratory for Radio Astronomy, Chinese Academy of Sciences, Nanjing 210008, China}
\altaffiltext{4}{Mizusawa VLBI Observatory, National Astronomical Observatory of Japan, Oshu 023-8501, Japan}
\altaffiltext{5}{Max-Planck-Institut f\"{u}r Radioastronomie, Auf dem H\"{u}gel 69, 53121 Bonn, Germany}
\altaffiltext{6}{Astron. Dept., King Abdulaziz University, P.O. Box 80203, Jeddah 21589, Saudi Arabia}


\begin{abstract}
Single point observations are presented in \nht~(1,1) and (2,2) inversion transitions using the Effelsberg 100 m telescope for a sample of 100 6.7 GHz methanol masers and mapping observations in the \coi\ and \coii\ $(1-0)$ transitions using the PMO Delingha 13.7 m telescope for 82 sample sources with detected ammonia. A further 62 sources were selected for either \coi\ or \coii\ line outflow identification, producing 45 outflow candidates, 29 using \coi\ and 16 using \coii\ data. Twenty-two of the outflow candidates were newly identified, and 23 had trigonometric parallax distances. Physical properties were derived from ammonia lines and CO outflow parameters calculated. Histograms and statistical correlations for ammonia, CO outflow parameters, and 6.7 GHz methanol maser luminosities are also presented. No significant correlation was found between ammonia and maser luminosity. However, weak correlations were found between outflow properties and maser luminosities, which may indicate that outflows are physically associated with 6.7 GHz masers.
\end{abstract}


%% subject header.  Objects should be in the appropriate "individual"
%% headers (e.g. quasars: individual, stars: individual, etc.) with the
%% additional provision that the total number of headers, including each
%% individual object, not exceed six.  The \objectname{} macro, and its
%% alias \object{}, is used to mark each object.  The macro takes the object
%% name as its primary argument.  This name will appear in the paper
%% and serve as the link's anchor in the electronic edition if the name
%% is recognized by the data centers.  The macro also takes an optional
%% argument in parentheses in cases where the data center identification
%% differs from what is to be printed in the paper.



\section{Introduction}           %% first-level sections will be auto-capitalized
\label{sect:intro}

%\textbf{MANUSCRIPT:}

Massive stars ($>8$M$_{\odot}$) play a dominant role in shaping galactic structure and evolution. Although \cite{Shu1987} provided a detailed description on low mass star formation, massive star formation remains under debate. \cite{Zinn2007} summarized the massive star formation process into a four-phase evolutionary sequence, similar to low mass stars, and proposed that high mass star formation is not merely a scaled up version of low mass formation, but involves new and different physical processes. However, challenges in observing high mass star forming regions has hindered understanding, with massive stars being statistically rare, at large distances, evolving rapidly with short lived evolutionary phases, deeply buried in dense molecular envelopes, and usually interacting with nearby complex star forming compounds \citep{Sheph1996a,Zinn2007}.

Signposts to trace massive star formation regions are usually water, methanol, and hydroxyl masers. Class II methanol masers in the $5_1-6_0 ~\mathrm{A}^+$ transition (at 6668.5192 MHz) \citep{Menten1991, Sobo1997} are brightest after H$_2$O and are suggested to be only associated with massive star formation. The 6.7 GHz methanol masers appear before the UC\HII~region phase (hot core phase), and disappear as the UC\HII~region evolves \citep[e.g.][]{Code2000,Code2004,van2005}. The 6.7 GHz methanol maser pumping mechanism is considered to be radiative, requiring specific temperatures and column densities that are not expected in low mass young stellar objects (YSOs) \citep{Cragg2005}. Observations of low mass star forming regions have supported this non-detection (e.g. \citealt{Minier2003}, \citealt{Bourke2005}, \citealt{Pan2008}, \citealt{Green2012iv}). \citet{Breen2013} also examined some masers associated with evolved stars and confirmed the 6.7 GHz methanol masers are exclusively associated with massive star forming regions.

Massive stars are considered to be initially born in giant molecular clouds (GMCs), so one might essentially want to know the physical conditions in the star forming regions. The inversion lines of ammonia, particularly the NH$_3$ (1,1), (2,2), and (3,3) lines, are excellent thermometers of the dense gas because they are collision excited and the molecules are not easily depleted onto dust grains \citep{Ho1983, Mang1992, Berg1997}. Therefore, key gas parameters, such as temperature, opacity, and column density can be estimated by measuring inversion line intensity. Physical property differences between weak and strong 6.7 GHz methanol masers have been discussed widely. \cite{Szym2000} suggested that IRAS colors vary with maser luminosity, but \cite{Pan2007ii} did not confirm this difference. \cite{Wu2010} found physical properties differences derived from ammonia lines and 6.7 GHz methanol maser luminosity, but \cite{Pan2012} later disproved these findings with a larger and less biased sample.

Molecular outflows are useful probes of star-forming activities in early phases. In the disk-jet model of low mass star formation, molecular outflows are a phenomenon of surrounding gas entrained by high-velocity jets or star winds. As the first CO outflow was discovered in Orion KL by \cite{Kwan1976}, accumulated observations of high mass star formation regions show outflows are also common in massive star formation regions \citep[e.g.][]{Snell1990, Sheph1996a, Ridge2001, Beu2002, Wu+2004, Xu2006, Arce2007, Vil2014I, Maud2015}, indicating a similar driven mechanism. Despite the similarity, whether there are differences between low and high mass star formation region outflows \citep{Beu2002,Wu+2004,Zinn2007} remain under debate.

Outflow properties are correlated with physical properties of the central source \citep{Beu2002, Wu+2004, Zhang2005, Vil2014I}. Methanol masers are also tracers of massive star formation, which leads to the question of whether outflows associate with methanol masers. \cite{Minier2000,Minier2001,Minier2002} and \cite{Code2004} found that H$_2$O and CH$_3$OH masers were closely associated with the evolutionary phase when outflows are present. \cite{Vil2014I,Vil2015II} analyzed CO $(J=3-2)$ line data in 54 6.7 GHz methanol maser sources and identified 44 resolvable methanol maser associated outflows (MMAOs). They also investigated relationships between outflow and maser properties, and suggested that the maser pumping source may be the outflow driver.

Whether the luminosity of a methanol maser indicates different physical conditions traced by ammonia remains to be verified. In this work, we expand the sample of methanol masers to check the conclusions of \cite{Pan2012}. Although research on massive outflows tends to focus on high resolution observations, it is still necessary to expand the number of known outflow candidates in massive star formation regions through single dish surveys. In this study, ammonia properties and outflow identification was investigated exclusively around 6.7 GHz methanol masers to verify and investigate potential relationships. Some of the 6.7 GHz maser samples have accurate parallax distances, which can provide more reliable physical properties. Previous single dish surveys of molecular outflows in massive star formation regions have focused on molecular transitions such as CO $(2-1)$ and $(3-2)$. The present survey searched for \coi~$(1-0)$ and \coii~$(1-0)$ outflows around 6.7 GHz methanol masers using the PMO 13.7 m telescope.

Section~\ref{sect:Obs} describes the sample, observations, and data reduction. Data analysis and derivation of physical properties are presented in Section~\ref{sect:data analysis}. Outflow detection frequency, and relationships among ammonia, CO outflow, and maser properties are discussed in Section~\ref{sect:discussions}. Section~\ref{sect:summary} summarizes and presents the main conclusions.

\section{Observations and Data Reduction}
\label{sect:Obs}

\subsection{Sample selection}
\label{sect:sample}

\begin{figure*}[!t]
\centering
\plotone{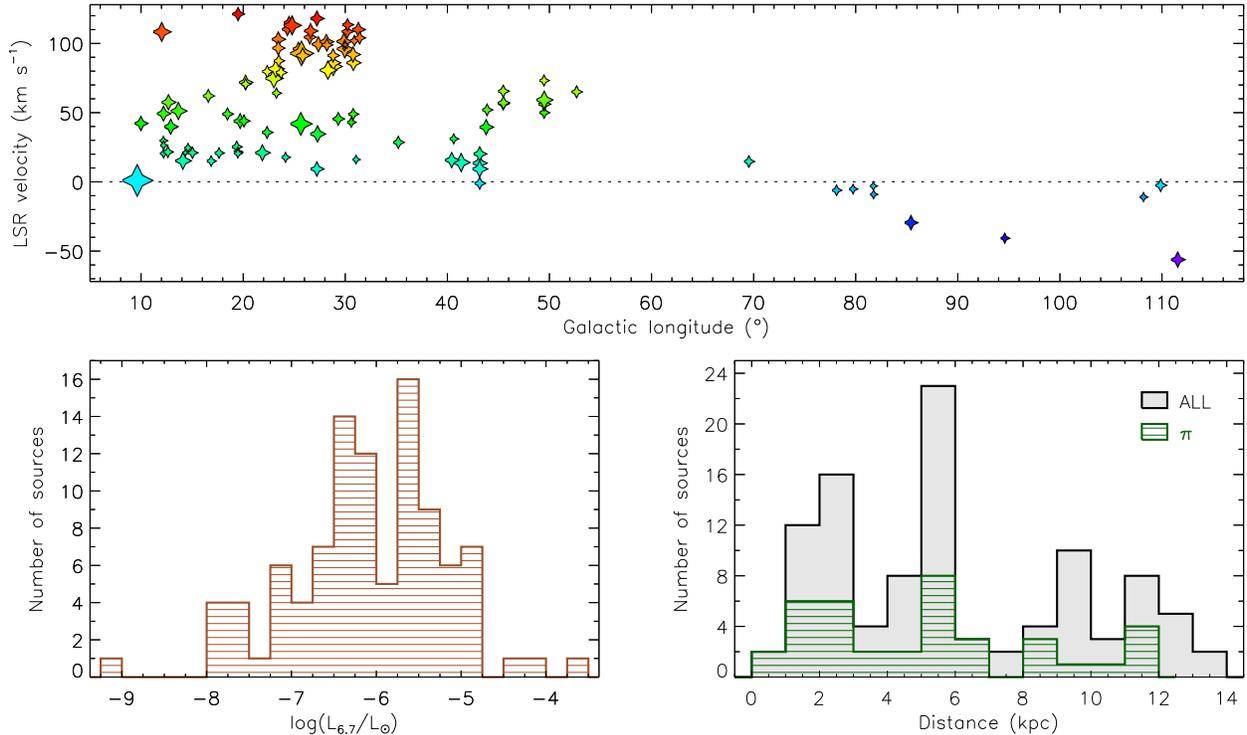}
\caption{Top: Sample locations. Filled stars show the maser position. Symbol size represents maser luminosity. Bottom left: Methanol maser luminosities (6.7 GHz). Bottom right: Sample distances, $\pi$ represents parallax distances.}\label{fig_gal_dist}
\end{figure*}

A sample of 100 sources with single point ammonia observations were selected from \cite{Cas2009} and \cite{Xu2009} with decl. $> -20^{\circ}$ and position accuracy better than 1$''$. he positions of these sources were determined by interferometer observations. Some sources have accurate trigonometric parallax distances calculated by the Bar and Spiral Structure Legacy (BeSSeL) survey and Japanese VLBI Exploration of Radio astronomy (VERA) (see \citealt{Reid2009i} and their serial papers; \citealt{Reid2014}). Distances of the remaining sources were determined kinematically from their observed radial velocities by applying model A5 in \cite{Reid2014} with a FORTRAN script by \cite{Reid2009vi}. Near or far distance ambiguities are either resolved using HI self-absorption or referring to allocations in the literature \citep{Green2011,Dun2011,Sch2011}. If a source could not be resolved using HI self-absorption and have no previous allocations in the literature, we then take the near distance for it. The number of such sources is less than 10\% of the sample.

\begin{figure}[!t]
\centering
%if two-column then
\epsscale{0.5}
%else comment it.
\plotone{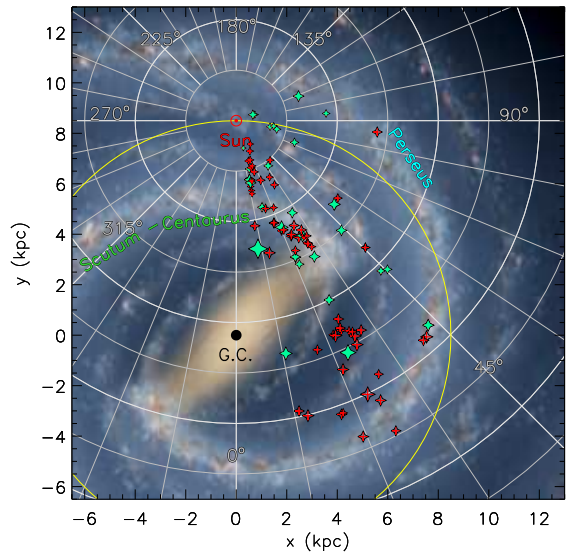}
\caption{Sample sources overlaid on Milky Way skeleton by R. Hurt. Filled stars represent sources with accurate parallax distances, open stars with red $+$ have kinematic distances.}\label{fig_maser_dist}
\end{figure}

Table~\ref{tab1} shows the chosen source maser properties, and their galactic locations are shown in the top panel of Fig.~\ref{fig_gal_dist}, with their distance distribution in the bottom-right panel. The sample covers a relatively wide range of maser luminosities (calculated from peak flux density, assuming isotropic emission and a typical line width to be 0.25 \kms) from $8.75 \times 10^{-10}$ to $2.36 \times 10^{-4}$ \Lsun, as shown in the bottom left panel of Fig.~\ref{fig_gal_dist}. Sample sources overlaid on an artist's conception of the Milky Way galaxy seen from far Galactic North (R. Hurt: NASA/JPL-Caltech/SSC) is shown in Fig.~\ref{fig_maser_dist}.

\subsection{Ammonia observations}
\label{sect:nh3 obs}

Single point observations of the sample in ammonia inversion transitions \nhti\ \nhtii\ were performed during March 2011 and May 2012 using the Effelsberg 100~m telescope. A dual channel cooled K-band HEMT receiver with two polarizations (LCP/RCP) was used as the frontend, and a fast Fourier transform spectrometer (FFTS) as the backend. The FFTS was set to Narrow-Band-IF mode with a bandwidth of 100~MHz at 24~GHz. It had 32,768 channels in each IF inputs, allowing simultaneous observation of the two lines, with velocity channel separation 0.038~ \kms. The observations were made in frequency switched mode with frequency throw 7.5~MHz. Integration times ranged from 4--13 minutes depending on the system temperature, which varied from 60--90~K. Pointing accuracy was found to be better than 10$''$. Flux calibration was accurate to 10\%, estimated by observing the standard source W3(OH). Flux density was converted to main beam brightness temperature (\Tmb), assuming a conversion factor of 1.36 K Jy$^{-1}$ \citep*[using Equation 8.20 in][]{Wilson2009}. Half power beam width (HPBW) of the telescope at the observed frequency was approximately 40$''$. All spectra were smoothed to a velocity channel separation of 0.32~\kms. \nht\ data were reduced using the CLASS package of the GILDAS\footnote{\url{http://www.iram.fr/IRAMFR/GILDAS/}} software distribution developed by IRAM.

\subsection{CO observations}
\label{sect:CO obs}

\coi, \coii\, and \coiii\ $J = 1 - 0$ data were obtained using the Purple Mountain Observatory (PMO) Delingha 13.7 m millimeter telescope in May and June 2014, and supplementary observations were performed in June 2015. A $10' \times 10'$ area was mapped around each sample source, using a 3$\times$3 multi-beam superconducting spectroscopic array receiver (SSAR) with a two sideband superconductor insulator superconductor (SIS) mixer as the frontend \citep{Shan2012}. The receiver enabled the three CO lines to be simultaneously observed, \coi\ line in the upper sideband (USB), and \coii\ and \coiii\ in the lower sideband (LSB). The backend employed a high definition FFTS with 1~GHz bandwidth. The spectrometer provided 16,384 channels, corresponding to velocity channel separation 0.16~\kms\ for \coi\ and 0.17~\kms\ for \coii\ and \coiii. Typical system temperatures were approximately 210~K for USB and 130~K for LSB measurements. Pointing accuracy and tracking errors were better than 5$''$. HPBW at 115.271~GHz was approximately 52$''$. Mapping observations were made using on the fly (OTF) mode with a scan speed of 50$''$ s$^{-1}$ and a step size of 15$''$ along Galactic longitude or latitude.
Standard sources were observed at intervals to estimate main beam efficiencies (\etamb), which were calibrated by incorporating an elevation based antenna gain curve\footnote{See the routine status report at \url{http://www.radioast.csdb.cn/zhuangtaibaogao.php}}. Mean \etamb\ were 48\% for USB and 52\% for LSB, with variations dominated by elevations. The antenna temperatures were converted to main beam temperatures using \etamb\ above. Mean rms noise level of the spectra was approximately 0.5	K for \coi\ and 0.3~K for \coii\ and \coiii\ after subtracting the linear baseline fit.
Raw data were processed using pipeline scripts written in CLASS and GREG software packages. Spectral data were then meshed into cubes with grid spacing 30$''$. All data were saved as astronomical FITS files for subsequent analysis.

\section{Data Analysis and Results}
\label{sect:data analysis}

\subsection{Ammonia line parameters}
\label{sect:nh3 obs para}

\begin{figure*}[!t]
\centering
\epsscale{1}
\plotone{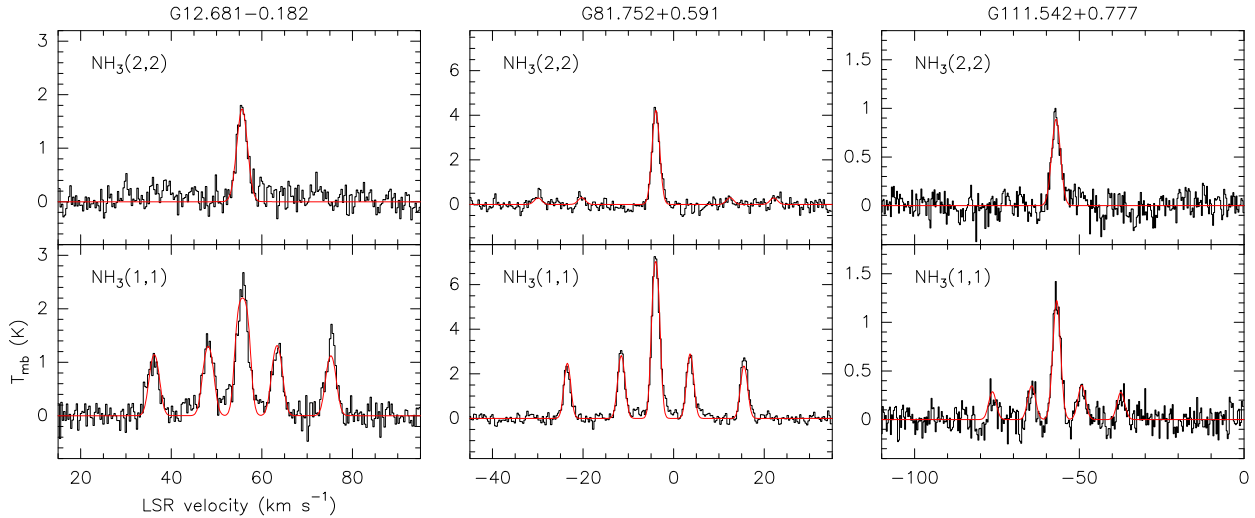}
\caption{Three sample source \nht\ spectra with fitting lines for illustration. The NH3(1,1) method was used in CLASS to estimate opacities. The GAUSS method was employed to calculate FWHM for lines with weak satellite components.}\label{fig_nh3_spec}
\end{figure*}

Ammonia inversion transitions were observed to estimate the dense gas temperatures around the methanol maser site. A total of 82 sources show \nhti\ and 73 \nhtii\ at the 3$\sigma$ level, 0.53 K. Fitting method NH3(1,1) was chosen for CLASS to model \nhti\ lines incorporating hyperfine structure as well as deriving opacities, with the assumptions of Gaussian velocity distribution and equal excitation temperature. For \nhtii\ spectra, which usually show weak hyperfine components, the main component was fitted using the GAUSS method. Relevant fitting parameters and uncertainties, such as radial velocities with respect to the local standard of rest, brightness temperatures, line widths, and optical depths, are shown in Table~\ref{tab2}. Figure~\ref{fig_nh3_spec} shows processed \nhti\ and \nhtii\ spectra for three characteristic sample sources.

\subsection{Ammonia line properties }
\label{sect:nh3 phy para}

Physical parameters of the dense environments around maser sites were derived from \nhti\ and \nhtii\ transitions, providing excitation, kinetic, and rotational temperatures, as well as column density. The methods are described in appendix~\ref{sect:A_nh3}, and the derived physical parameters are shown in Table~\ref{tab3}.

\begin{figure*}[!t]
\centering
\plotone{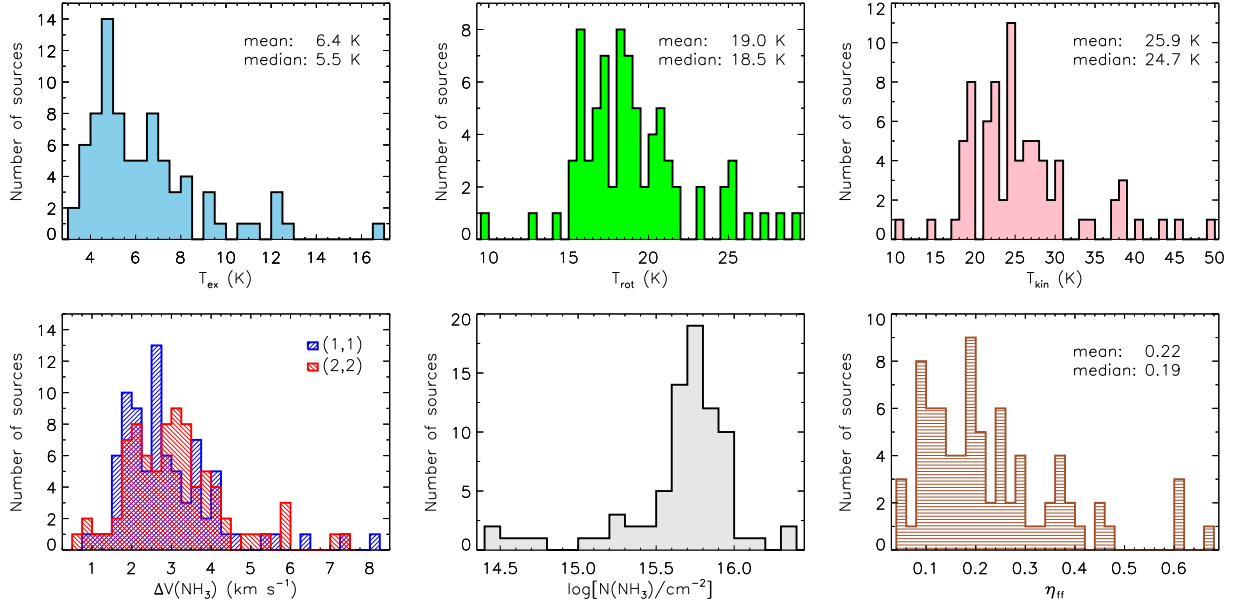}
\caption{ \nht\ line properties. Top three panels (left to right): excitation temperature, rotational temperature, and kinetic temperature. Bottom three panels (left to right): FWHM, \nht\ column density, and filling factor.}\label{fig_nh3_hist}
\end{figure*}

Figure~\ref{fig_nh3_hist} shows the statistics of observed and physical properties for \nht\ detect samples. The left top panel shows the excitation temperature, \Tex, which ranges approximately 3.0--13.0~K, with mean and median 6.4~K and 5.5~K, respectively. These are lower than those of the low luminosity 6.7 GHz methanol maser sites in \cite{Wu2010}, similar to those of the high luminosity 6.7 GHz methanol masers in \cite{Wu2010}, but slightly higher than those studied by \cite{Pan2012}. These low values are most likely due to the small beam filling factor, $\eta_{\mathrm{ff}}$. If \Tex\ = \Trot, then the typical beam filling factor turns = 0.19, significantly larger than 0.07, derived by \cite{Pan2012}.

The top middle panel of Figure~\ref{fig_nh3_hist} shows the sample rotational temperatures, which have a mean and median of 19.0~K and 18.5~K, respectively, slightly lower than those of \cite{Wu2010}. Mean and median kinetic temperatures are 25.9~K and 24.7~K, similar to \cite{Pan2012} and lower than \cite{Wu2010}. The left bottom panel of Fig.~\ref{fig_nh3_hist} shows \nhti\ and \nhtii\ FWHM for main component emission, with typical values 2.7 and 3.0~\kms, respectively. The line widths are comparable with \cite{Wu2010} and \cite{Pan2012}. And the ratio of the typical \nhtii\ to \nhti\ line width is 1.1. Typical \nht\ column density of the sample masers was ${5.4\times10^{15}}$ cm$^{-2}$, similar to \cite{Pan2012}.

%\placefigure{fig1}

\subsection{CO outflow identification}
\label{sect:out id}

Line profiles of \coi$(1-0)$ and \coii$(1-0)$ spectra at maser sites of the 82 sources with \nht~detection were examined, and 62 were considered to be suitable for outflow identification. The selection criterion was that either \coi\ or \coii\ line wings should not be contaminated by velocity components along the line of sight that are not associated with 6.7~GHz maser emission. Both \coi\ and \coii\ spectra were used for outflow identification. Outflow wings are expected to be most obvious in \coi$(1-0)$ due to its large abundance, but \coi$(1-0)$ is often contaminated by extra components, while for \coii$(1-0)$, which has a lower abundance than \coi$(1-0)$, can be used in those cases where the wings of \coi$(1-0)$ have contamination.

\begin{figure*}
%\epsscale{.80}
\centering
\plotone{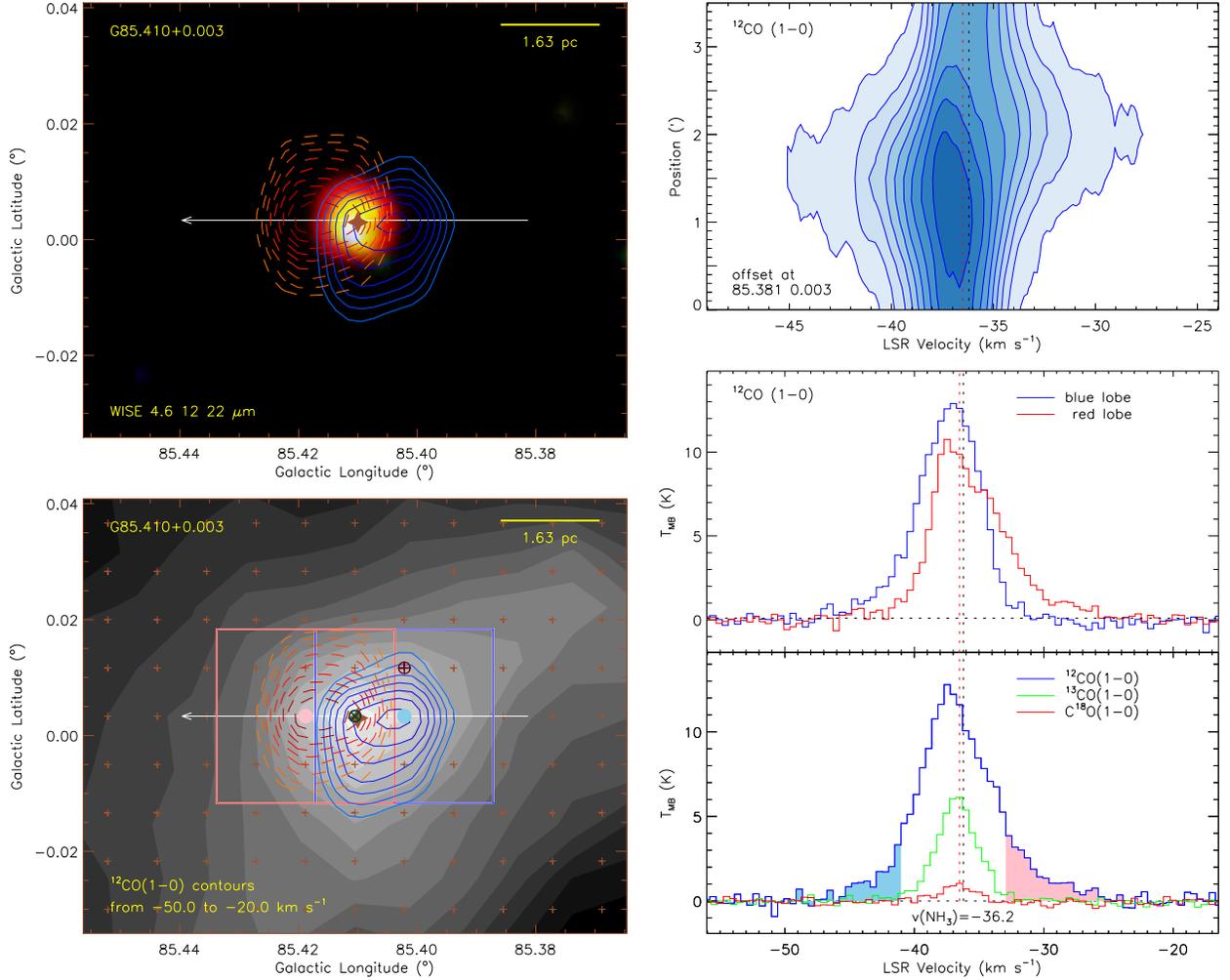}
\caption{%\small
Sample outflow identification for the G85.410+0.003 methanol maser.
The bottom left panel shows the outflow contours overlaid on integrated intensity of \coi\ depicted as filled contours, which has LSR velocity range of -50 -- -20~\kms\ with contours from 10~K~\kms\ at step 12~K~\kms. Brown crosses indicate CO data pixels.
The maser position is marked with a brown filled star, and the yellow scale line in each panel is 1~arcmin. CO emission peaks are marked on the bottom left panel with dark blue and green circles around $\times$ and dark red circles around $+$ for \coi, \coii\, and \coiii, respectively.
Then the lines shown in the bottom right panel are feature spectra extracted from the position of peak emission in the CO map. Red dashed vertical line is peak \coii\ velocity, and black dashed vertical line is \nht\ velocity for comparison. Color filled areas under the line are the selected line wings.
Integrating along the outflow wing ranges, we obtain the outflow lobes, which are illustrated as the blue and red contours within the blue and red boxes in the bottom left panel. The outflow contours start from 3.92 and 4.21 K~\kms\ with step 0.78 and 1.12~K~\kms\, respectively, to highlight prominent features. The spectra at the two lobe peaks are shown in the middle right panel. Blue and red lobe peak positions are marked in the bottom left panel with filled sky blue and pink circles, respectively. A position-velocity (P-V) slice diagram along the white arrow in the bottom left panel is shown in the top right panel. And finally a WISE 4.6, 12 and 22 $\mu$m false color image overlaid by outflow contours is shown in the top left panel.}\label{fig_out_sample}
%Top and bottom left panels: outflow contours overlaid on a WISE 4.6, 12 and 22 $\mu$m false color image (top left panel) and integrated intensity of \coi\ depicted as filled contours (bottom left panel). \coi\ integrated intensity has LSR velocity -50 -- -20~\kms, with contours from 10~K~\kms\ at step 12~K~\kms. Blue and red outflow contours start from 3.92 and 4.21 K~\kms\ with step 0.78 and 1.12~K~\kms\, respectively, to highlight prominent features.
%The maser position is marked with a brown filled star, and the yellow scale line in each panel is 1~arcmin. CO emission peaks are marked on the bottom left panel with dark blue and green circles around $\times$ and dark red circles around $+$ for \coi, \coii\, and \coiii, respectively. Brown crosses indicate CO data pixels. Blue and red boxes show boundaries of blue and red lobe contours.
%Bottom right panel: \coi, \coii\, and \coiii\ spectra at emission peak positions. Red dashed vertical line is peak \coii\ velocity, and black dashed vertical line is \nht\ velocity for comparison. Color filled areas under the line are the selected line wings.
%Middle right panel: spectra at lobe peaks. Blue and red lobe peak positions are marked in the bottom left panel with filled sky blue and pink circles, respectively.
%Top right panel: position-velocity (P-V) diagram along the white arrow in the top and bottom left panels.}\label{fig_out_sample}
\end{figure*}

The following procedure was used to identify outflow candidates, with an example case, G85.410+0.003, shown in Figure~\ref{fig_out_sample}.

\textit{Find peaks and extract feature spectra.}
The CO emission peak emission positions were located on the integrated intensity images, as shown in the bottom left panel of Figure~\ref{fig_out_sample}. Although the \coi\ and \coii\ peak positions for G85.410+0.003 coincide with each other, the \coiii\ peak is offset one pixel ($\sim 1.1$ pc), probably due to noise ambiguity and grid seperation. Generally, the \coi, \coii\ and \coiii\ peaks do not coincide with each other and the maser point. Some peaks have offsets of one to two pixel spacing ($\sim 30''$). When sometimes the \coi, \coii\ and \coiii\ peaks are all different from each other, then the position of either \coi\ or \coii\ peak that closer to the maser point was chosen to extract feature spectra.
Spectra were then extracted from data cubes at the chosen peak position, providing feature lines for outflow identification (bottom right panel of Figure~\ref{fig_out_sample}).
To identify high velocity outflow components, the extracted spectra were smoothed to velocity resolution of 0.5--1.0~\kms\ to reduce noise in an individual channel. The core velocities (\vpeak) were obtained at the line intensity peak for either \coii\ or \coiii\, corresponding to the nearest maser and \nht\ peak velocity. Peak \coii\ velocity was chosen when \coiii\ emission was weak.

\textit{Outflow wings.}
Outflow wing ranges ($\Delta V_{\mathrm{b/r}}$) were determined by line diagnosis. Outflow wing is the component left after subtracting the Gaussian core profile. Since \coii\ and \coiii\ trace the denser core gas around YSOs, wing ranges were determined by comparison with different lines \citep{Lada1985}.
For a typical \coi\ line, blue and red wing velocity ranges were determined by \coi\ velocity extent, where the \coii\ line has no emission (e.g. the filled area of the line in the bottom right panel of Figure~\ref{fig_out_sample}). \coii\ emission was treated as the inner core component. Note that wing ranges were also manually adjusted according to the position velocity (P-V) diagram in the top right panel of Figure~\ref{fig_out_sample}, which is described in the next procedure. The P-V diagram was included in Figure~\ref{fig_out_sample} to illustrate the velocity structure along the slice crossing outflow lobes.
\coii\ line wings were obtained similarly, but compared with \coiii\ lines. \coi\ wings are generally high velocity outer wings, while \coii\ wings indicate inner wings, as discussed in \cite{Lada1985}.

\textit{Outflow contours and P-V diagrams.}
Outflow image was obtained by integrating the intensity along the wing velocity range (e.g. the top left and bottom left panels of Figure~\ref{fig_out_sample}).
Integrated contours were presented to highlight prominent features, starting mostly from 40\% to 70\% of the peak integrated intensities.
The contours were then checked by naked eyes if they can be regarded as an outflow lobe. A blue or red box was added to restrict the boundary of a lobe (e.g. the blue and red boxes on the bottom left panel in Figure~\ref{fig_out_sample}). Lobe area ($A_{\mathrm{b/r}}$) was calculated based on contours.
Peak positions of lobes were located on the outflow contours (e.g. the blue and red filled circle on the bottom left panel in Figure~\ref{fig_out_sample})), and the spectra at lobe peaks were extracted for comparison (e.g. the middle right panel of Figure~\ref{fig_out_sample}).
Lobe length ($l_{\mathrm{b/r}}$) was then measured generally from the maser postion to the furthest radial distance along the line that connects the positions of two lobe peaks or one lobe peak to the central maser position (e.g. the white arrow shown in the bottom left panel of Figure~\ref{fig_out_sample}). If outflow lobe peaks and maser point appeared to be on top of each other, the lobe length was then measured along the major axis.
A P-V slice image (e.g. the P-V image on the top right panel in Figure~\ref{fig_out_sample}) was also extracted along the line. For a typical outflow, the P-V diagram should show velocity bulges comparing non-outflow positions, and these velocity bulges were regarded as an evidence of outflow.

\textit{Comparisons with WISE false-color images.}
To uncover potential driving sources of the outflows, we used the high sensitivity mid-infrared images taken by the Wide-field Infrared Survey Explorer (WISE) \citep{Wright2010}. WISE maps the entire sky in four infrared bands W1, W2, W3, and W4 centered at 3.4, 4.6, 12, and 22 $\mu m$. A RGB false color image of WISE data at 4.6, 12, and 22 $\mu m$ overlaid with outflow contours was plotted for comparison. For instance, the infrared emission relating to the maser site can be clearly resolved in the top left panel of Figure~\ref{fig_out_sample}.
Most masers have a corresponding WISE detection at the same position.
However, a few masers are difficult to cross identify the WISE detection, but some seem to have offsetting WISE sources. As \cite{Vil2014I} stated, since there exist different hypotheses for the 6.7 GHz maser formation (they are either embedded in circumstellar regions around protostars or just in outflows). It is thus likely that some masers could be offset, while for a few other cases where the offset is too far, they probably do not have detectable WISE emission due to infrared extinction.

\begin{figure*}[!tp]
\centering
\plotone{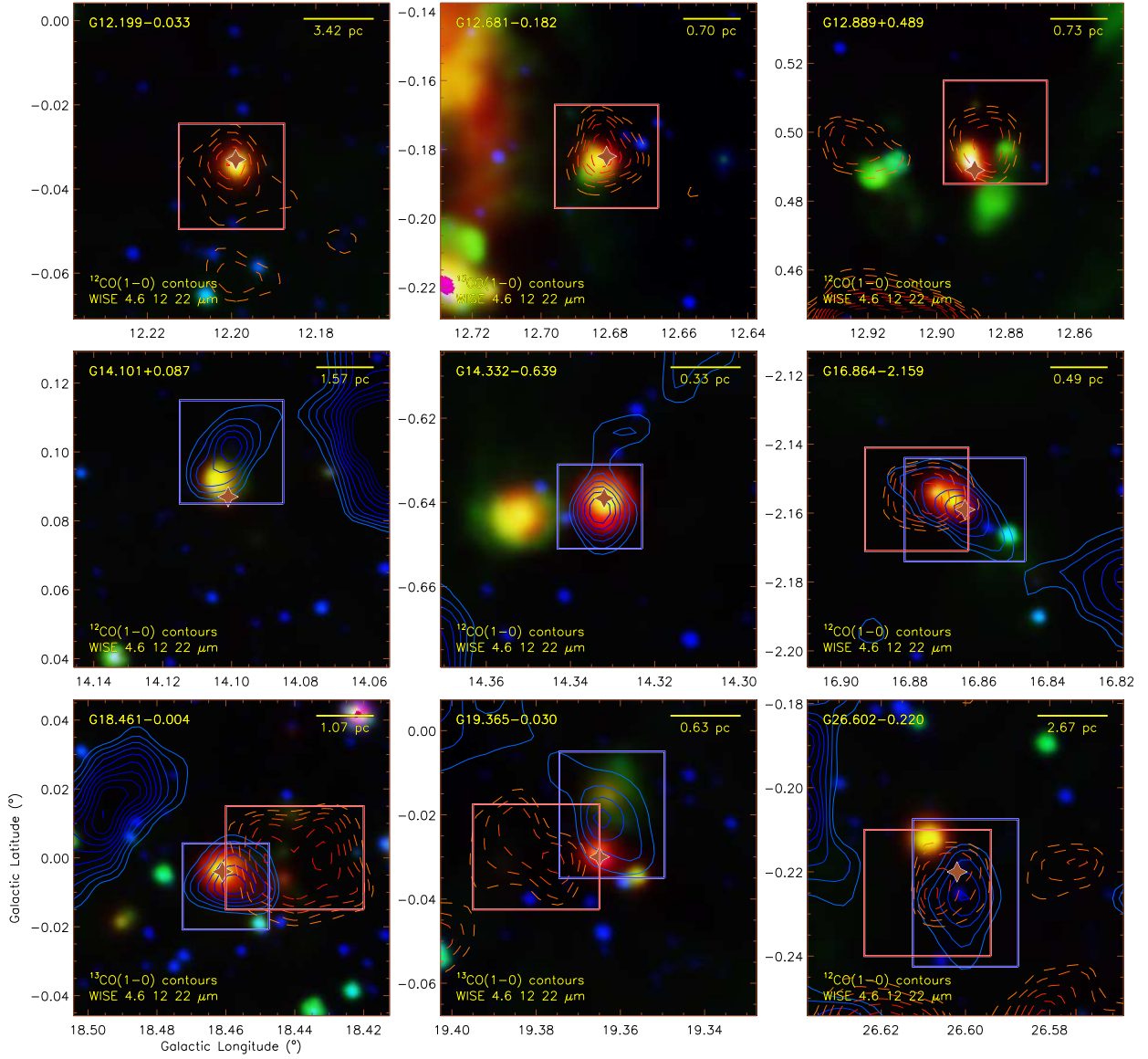}
\caption{Newly identified outflows. Outflow contour lower limits and steps (within the boxes) are presented to highlight prominent features. Contours are overlaid on corresponding WISE false color images. \coi\ and \coii\ lines selected for outflow identification are labelled in the bottom left corner of each image. Symbols within each panel are the same as Figure~\ref{fig_out_sample}.}\label{fig_new_out}
\end{figure*}

%\addtocounter{figure}{-1}
\begin{figure*}[!t]
%\epsscale{.80}
\centering
\figurenum{6}
\plotone{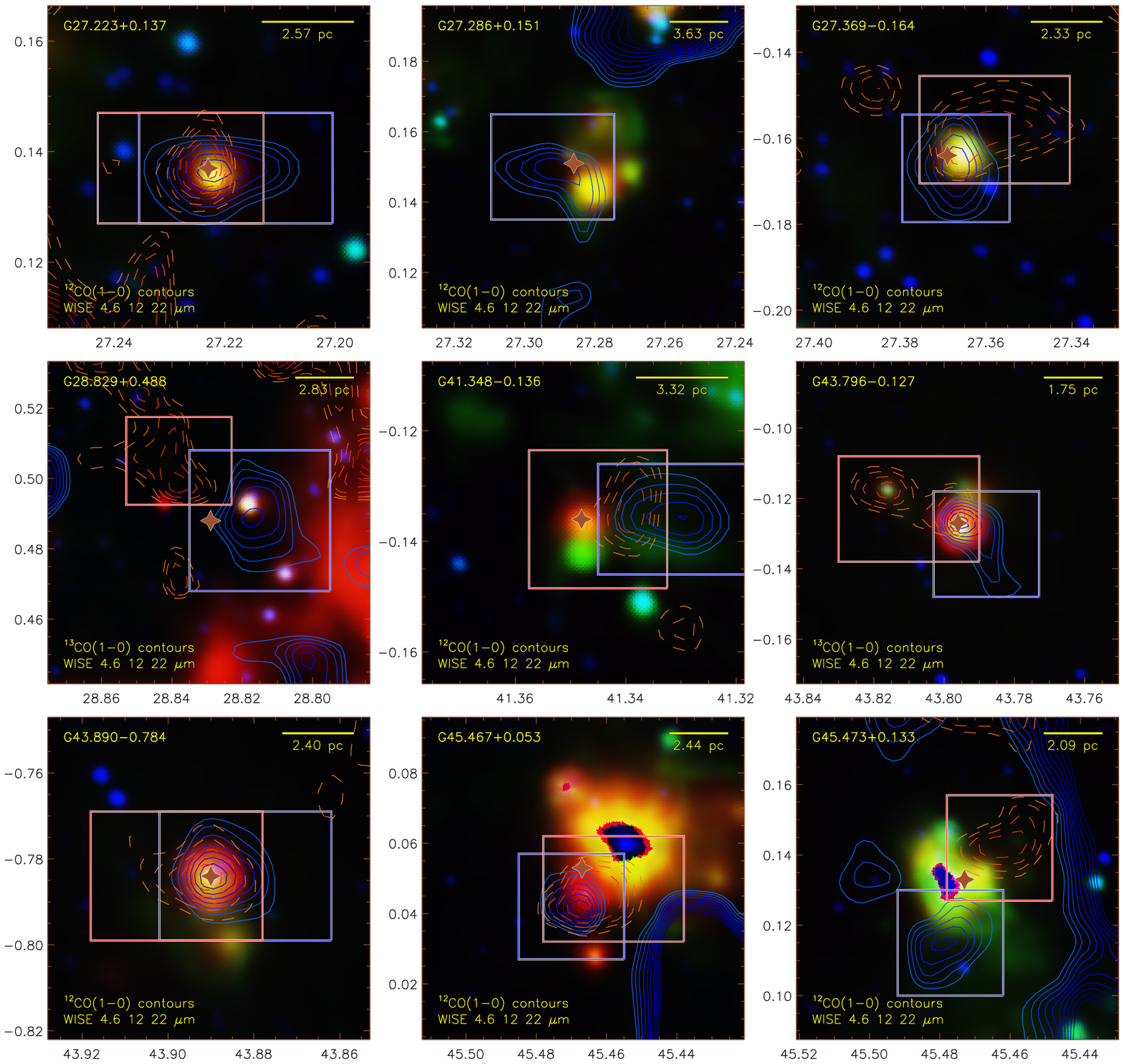}
\caption{Continued.}
%\\
%Figure 5.-- \textit{Continued}.
\end{figure*}

\begin{figure*}[!t]
%\epsscale{.80}
\centering
\figurenum{6}
\plotone{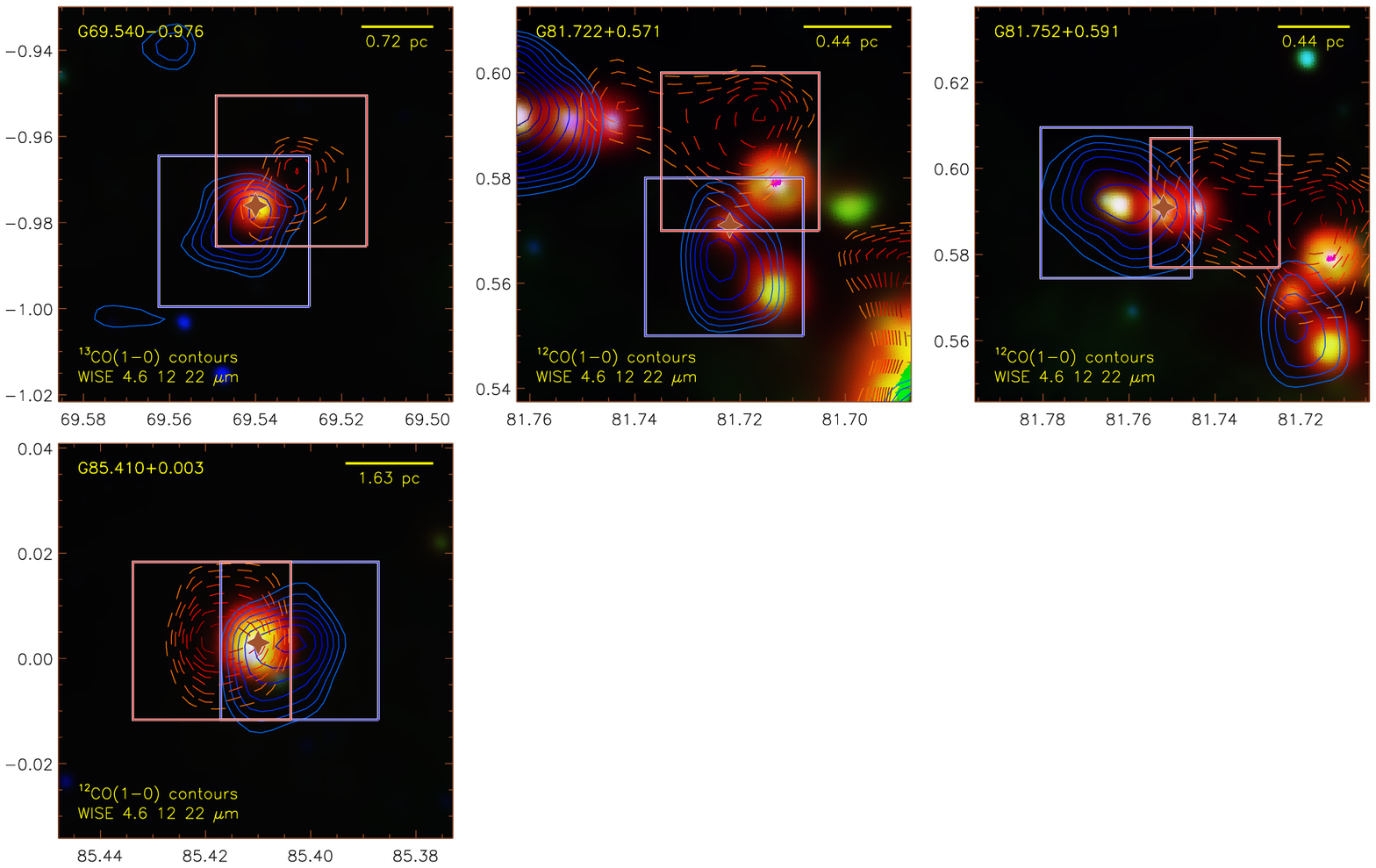}
\caption{Continued.}
\end{figure*}

Forty-five sources were identified to have significant CO outflow features from the 62 candidates with either \coi\ or \coii\ line wing. The detection frequency is 73\%. Among these, 22 were newly identified, 29 were diagnosed with \coi\ line wings and 16 with \coii\ line wings. Detailed discussion regarding outflow detection frequency is presented in Section~\ref{sect:dete rate}. The parameters for calculating outflow physical properties are shown in Table~\ref{tab4}, and the 22 newly identified outflows are overlaid on corresponding WISE false color images in Figure~\ref{fig_new_out}.

\subsection{CO outflow physical properties}
\label{sect:phy out}

Physical properties such as mass, energy, and momentum flux, must be estimated to allow outflow investigation. The estimates depend on estimates of optical depth, excitation temperature, and filling factor of the emitting gas, which provide column density as a function of velocity. The calculations are described in appendix~\ref{sect:B_co}. Among the 45 outflow candidates, 23 have trigonometric parallax distances.

\begin{figure*}[!t]
%\epsscale{.80}
\centering
\plotone{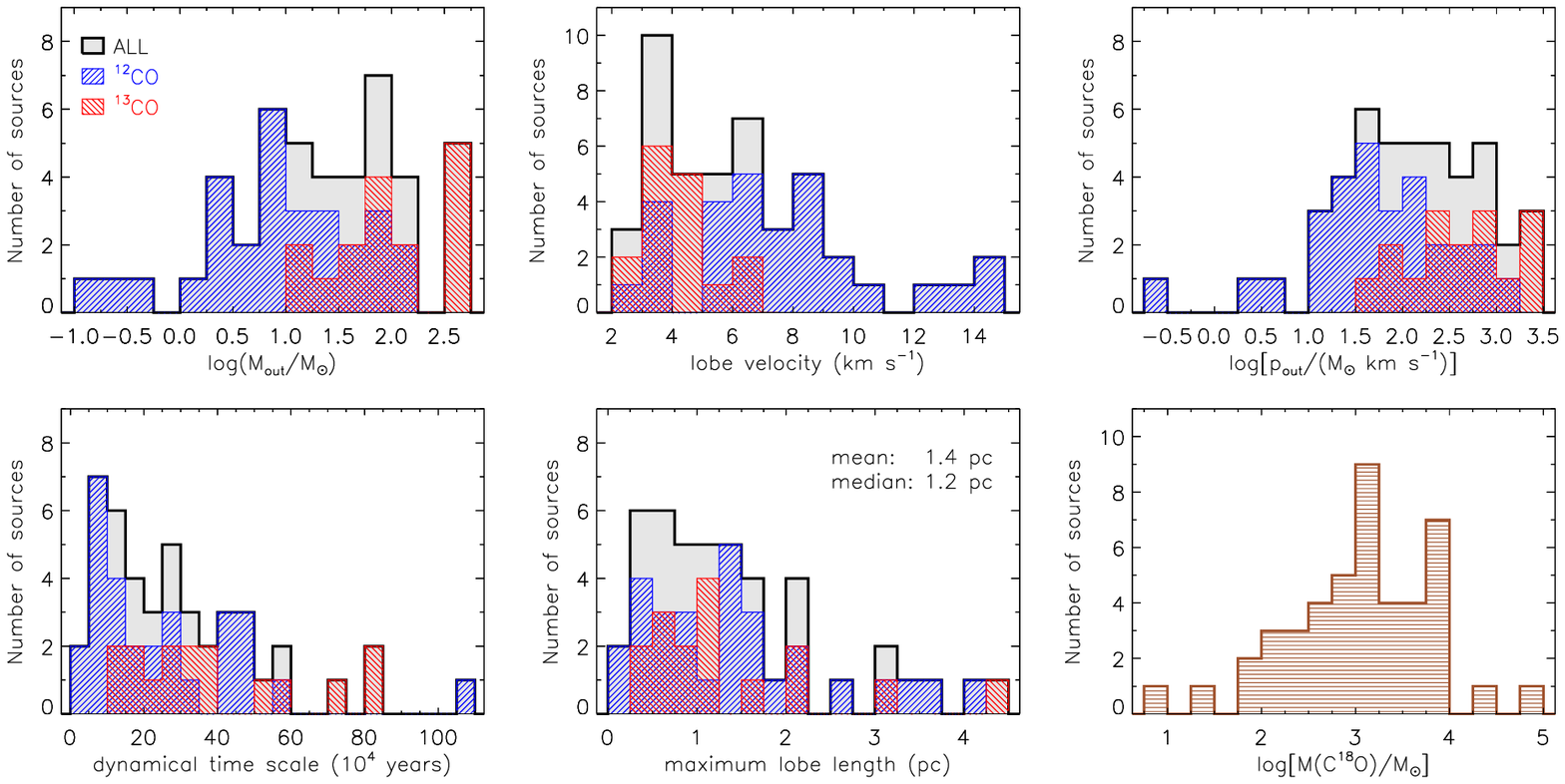}
\caption{CO outflow properties without inclination correction. \coi\ and \coii\ outflow properties are distinguished for comparison. Top three panels: outflow mass, average lobe velocity, and momentum, respectively. Bottom three panels: dynamical time scale and maximum lobe size, respectively, of outflows and \coiii\ cloud core masses.}\label{fig_co_hist}
\end{figure*}

The derived outflow physical properties are shown in Table~\ref{tab5}. However, it is difficult to determine the outflows' parameters precisely due to inclination, opacity and blending \citep{Arce2001}. Since \coii\ outflows are generally low velocity gas blended with ambient clouds, outflow mass may be overestimated. Inclination is also difficult to determine but has non-negligible effect on calculating outflow properties. Thus, statistically typical values were chosen for comparison with other works. Assuming a random distribution of inclination, the mean inclination was 57.3$^{\circ}$ \citep{Bon1996}. This implies lobe velocity should be scaled up by a factor of 1.9, while dynamical age should be reduced by a factor of 0.6.

Outflow properties are shown in Figure~\ref{fig_co_hist}, with typical \coi\ and \coii\ outflow masses 8 and 100~\Msun, respectively. The mass difference arises from different estimation methods and distances. For \coi, since the wing range does not show \coii\ emission, the line wings were assumed to be optically thin, leading to a lower limit mass estimation. The mean and median mass of \coiii\ cloud cores are both $1.3\times10^{3}$ \Msun. After inclination correction, the average lobe velocities of the sample outflows are typically 13~\kms\ and 7.5~\kms\ for \coi\ and \coii, respectively, because \coi\ line widths are generally wider than \coii\ and \coi/\coii\ velocity ratio is approximately 2 \citep[e.g.][]{Cabr1988,Sheph1996b,Nara2012}. In contrast to \cite{Vil2014I}, the wing velocities were not scaled for \coii\ outflows.

Momentum is shown in the top right panel of Figure~\ref{fig_co_hist}. Typical values corrected for inclination are 95 and 750~\Msun~\kms\ for \coi\ and \coii\ outflows, respectively. The mean maximum lobe length is 1.4~pc, incorporating the spatial resolution of the telescope and distance values. Outflow lobe sizes lower than the beam size limit ($\sim 55''$) were not able to be resolved. For the dynamical time scale presented in the bottom-left panel, typical values corrected for inclination are $1.1 \times 10^5$ and $2.0 \times 10^5$ years for \coi\ and \coii\ outflows, respectively. Such large values are an overestimation from the method described in Equation~(\ref{equ_t}). Since the $t_{\mathrm{d}} = l/v$ method can overestimate the flow age, \cite{Down2007} suggest a more accurate representation using $t_{\mathrm{d}} = 1/3~l_{\mathrm{lobe}}/\langle v \rangle$, which provides typical dynamical timescales $3.8 \times 10^4$ and $6.7 \times 10^4$ years for \coi\ and \coii\, respectively.

\section{Discussion}
\label{sect:discussions}

A statistical analysis of the physical properties of the 6.7 GHz methanol masers is performed to investigate the underlying physics. There are several caveats regarding the results and correlations.
\begin{itemize}
\item Sample selection. The sources in the Galactic range of $0^{\circ} < l < 20^{\circ}$ generally have multiple cloud components and distance ambiguities, adding uncertainty.
\item The chosen sample set did not include all the masers in \cite{Cas2009} and \cite{Xu2009} with declination greater than $-20^{\circ}$ due to limited observation time. A more uniform sample with more sources is required to increase the reliability of the conclusions.
\item The limited resolution and sensitivity of CO observations have large effects on determining outflow parameters.
\end{itemize}
Of the 45 identified outflow candidates, 23 have trigonometric parallax distances, and are treated separately in the statistical analysis.
However, the statistical correlations will assist better understanding of massive star formation processes.

\subsection{Outflow detection frequency}
\label{sect:dete rate}

The chosen samples were all 6.7 GHz methanol masers, which are exclusively associated with massive star formation. The outflow detection frequency for the methanol masers was 73\%. Of the 45 identified outflows, 29 have resolvable bipolar lobes. Those with one single resolvable lobe may be explained by their complicated environment. This detection rate only provides for a statistical valid result for CO observations using the PMO DLH 13.7 m telescope.
Some sources without identified outflow, e.g. G29.956-0.016, could be identified outflows with higher spatial resolution or better intensity at higher transitions of CO \citep{Vil2014I}. The high detection rate in previous outflow studies from massive YSOs shows that molecular outflows are a common phenomenon of massive star formation. Particularly when associating outflow with 6.7 GHz masers, the detection rate increases up to one hundred percent \citep{Vil2014I}.

\subsection{Outflow properties and source luminosity}
\label{sect:out ir}

\cite{Maud2015} selected a distance limited sample of 99 RMS MYSOs and identified 85 outflows. They showed that all outflow parameters scale with source luminosity and proposed two interpretations on the scaling relationships. One view was that the outflows are driven by the massive protostellar cores in an embedded cluster, supporting a scaled up version of the outflow driving mechanism verified for low mass stars. An alternative interpretation was that the relationships are observational effects, since outflow parameters are determined mostly by the entrained mass originating from the core, which allows for different star formation processes.

\begin{figure*}[!t]
\epsscale{1.0}
\centering
\plotone{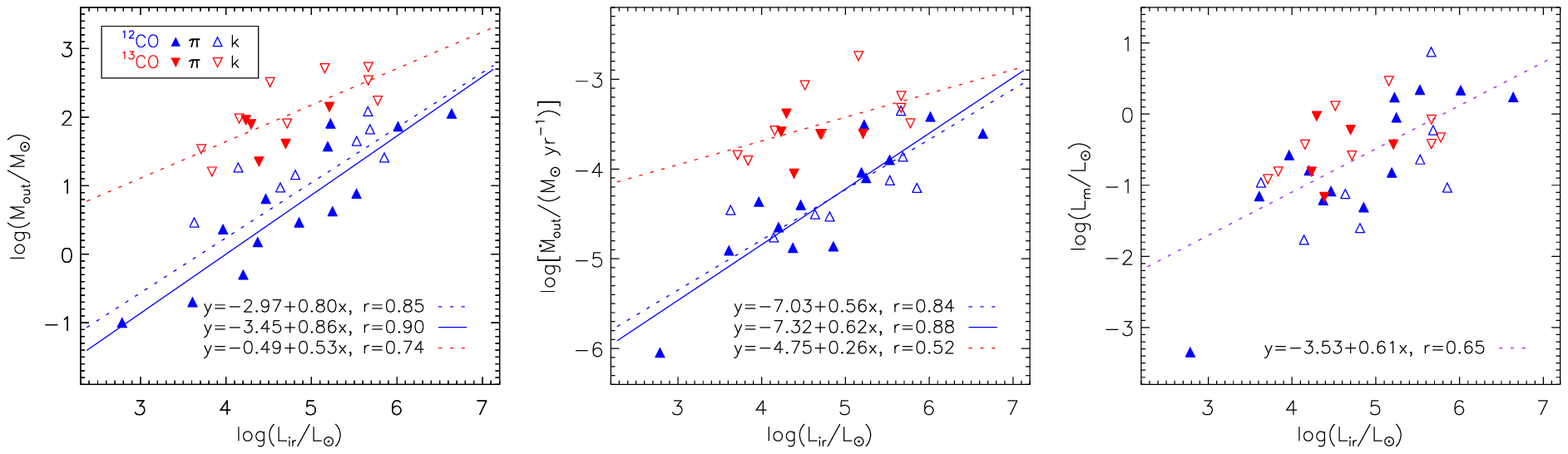}
\caption{CO outflow properties and IRAS infrared luminosity. Left panel: Outflow masses as functions of infrared luminosity. Blue and red solid lines are linear fittings for sources with parallax distances. Blue and red dashed lines are the fitted \coi\ and \coii\ outflows, respectively. $\pi$ and k represent source distances calculated using parallax and kinematics, respectively. Middle panel: As the left panel, but for outflow mass rate and infrared luminosity. Right panel: Mechanical luminosity of outflow as a function of infrared luminosity. Purple dashed line shows the fitted outflow.}\label{fig_out_ir}
\end{figure*}

The correlations of outflow properties and source luminosity were checked across the chosen sample. IRAS luminosities were used, since IRAS point catalog provides relatively reliable flux densities at 12, 25, 60, and 100~$\mu$m. \cite{Zhang2005} examined the relationships between outflow mass rate and mechanical force against IRAS far-infrared luminosity, and found no significant correlations. Outflow properties and IRAS infrared luminosity are shown in Figure~\ref{fig_out_ir}.

Tight correlations were found between outflow masses and infrared luminosity, with Pearson correlation coefficients $r=0.85$ and $r=0.74$ for \coi\ and \coii\ outflows, respectively. Since half of the outflow candidates have trigonometric parallax distances, these sources were fitted separately, with correlation coefficient $r=0.90$ for \coi\ outflows (the number of \coii\ outflows was too small to fit separately).

The mass outflow rate is also related to infrared luminosity, with $r=0.84$ and $r=0.52$ for \coi\ and \coii\ outflows, respectively. Again, when fitting \coi\ outflows with parallax distances, the correlation coefficient increased to $r=0.88$, although this is partly due to the reduction of the number of sources. The correlation coefficient for outflow mechanical luminosity and infrared luminosity was 0.65. While \cite{Zhang2005} found no apparent correlation between luminosity, mass outflow rate and mechanical force, for the current sample, these outflow properties scale with IRAS infrared luminosity.

\subsection{Ammonia line and outflow properties}
\label{sect:out nh3}

As discussed in Section~\ref{sect:phy out}, the typical dynamical timescale for 6.7 GHz methanol maser associated outflows approaches $10^5$ years, through to the development of UC\HII\ regions \citep{Wood1989}. The masers signpost a period of massive star formation before the UC\HII\ phase \citep{van2005}, while molecular outflows are also a phenomenon in the early phases. \cite{Code2004} provided a version of the evolutionary sequence in the hot core phase, and proposed: maser emission occurs before outflow; outflow appears and becomes detectable while maser emission is still present; maser emission disappears while outflow remains until the UC\HII\ region is formed. \cite{Vil2015II} amended the sequence and proposed that outflow appears and develops before maser emission. Maser emission appears when the in-fall process heats the dust grains and incubates certain conditions that can pump the methanol maser. Unfortunately, the dynamical timescales estimated by outflow surveys using single dish telescopes are too approximate to resolve the above evolutionary sequence. As noted in \cite{Maud2015}, interferometric observations are required to resolve this issue. However, since outflow phase encompasses the methanol maser phase and the dynamical timescale can be derived through the outflow, physical conditions and dynamical properties around maser sites can be checked for evolutionary trends.

\begin{figure*}[!t]
\centering
\plotone{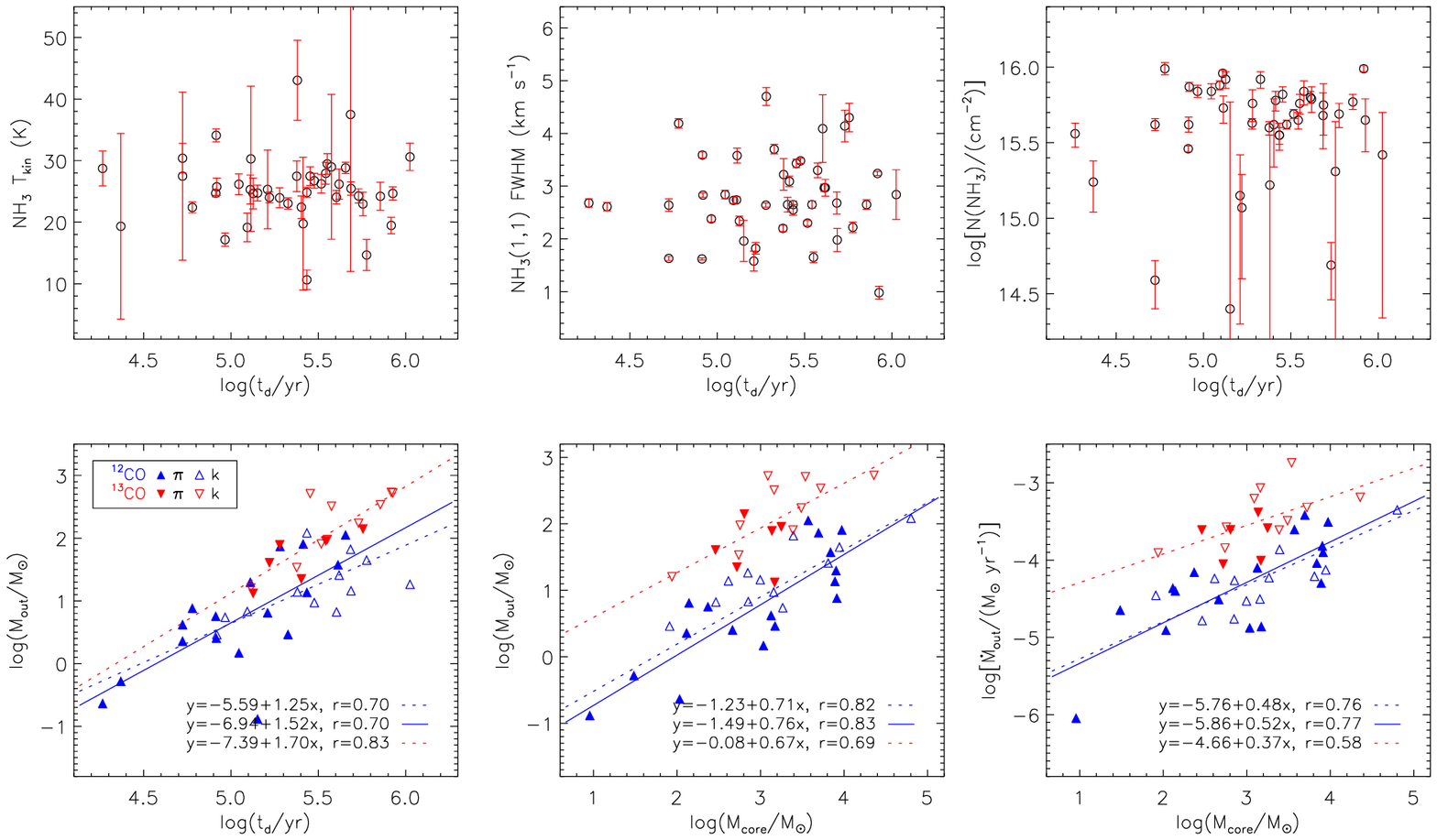}
\caption{\nht\ and CO outflow properties. Top three panels: \nht\ kinematic temperatures, line widths, and \nht\ column densities, respectively, as functions of outflow dynamical timescales. Bottom left panel:  Plot of outflow mass as a function of dynamical timescale. Bottom middle and right panels: outflow mass and outflow mass rate, respectively, as functions of core mass derived from \coiii\ emission. Blue and red dashed lines are linear fits for \coi\ and \coii\ outflows, respectively. Blue solid lines are linear fits for \coi\ sources with accurate parallax distances.}\label{fig_nh3_out}
\end{figure*}

The top three panels in Figure~\ref{fig_nh3_out} show the relations of \nht\ kinematic temperatures, line widths, and \nht\ column densities as functions of dynamical time. No significant correlations are present, which implies a lack of evolutionary trends.

The plot of outflow mass as a function of dynamical time, bottom left panel of Figure~\ref{fig_nh3_out}, shows significant correlations, with outflow mass increasing with dynamical timescale, correlation coefficients 0.70 and 0.83 for \coi\ and \coii\ outflows, respectively. Since both outflow mass and dynamical timescale relate to distance, such correlations can be expected. However, \cite{Vil2014I} failed to find this correlation, and argued that their range of dynamical times was probably too small. Uncertainties in determining outflow properties may also mask the correlations.

Many other works have checked the relationships of outflow masses and core or clump masses \citep[e.g.][]{Beu2002,Vil2014I,Maud2015}. The bottom middle panel in Figure~\ref{fig_nh3_out} shows the relation of outflow mass and the source \coiii\ core mass, with correlation coefficients 0.82 and $0.69$ for \coi\ and \coii\ outflows, respectively.  The bottom right panel of Figure~\ref{fig_nh3_out} shows the correlation of mass outflow rate and core mass, $r=0.76$, for \coi\ outflows. These correlations agree with \cite{Vil2014I}, and are similar to those discussed in \cite{Maud2015}. However, it is difficult to resolve whether the outflows arise from individual stars or multiple protostellar cores.

\subsection{Physical properties and maser luminosity}
\label{sect:maser relations}

Many previous surveys of 6.7 GHz methanol maser sources have investigated the correlation of maser luminosity and clump properties \citep[e.g.][]{Urq2013, Sun2014}. Though more massive protostars usually have larger maser luminosities, the difference of physical conditions between the high  and low luminosity masers remains undistinguishable. \cite{Wu2010} checked ammonia properties between high  and low luminosity 6.7 GHz methanol masers. Although several differences were found, uncertainties due to lack of sample sources weakened the results. \cite{Pan2012} expanded to a sample of 77 masers with single point ammonia observations. Aside from a weak correlation between maser luminosity and line width, no significant correlations were observed between the physical properties and maser luminosity.

\begin{figure*}[!t]
\centering
\plotone{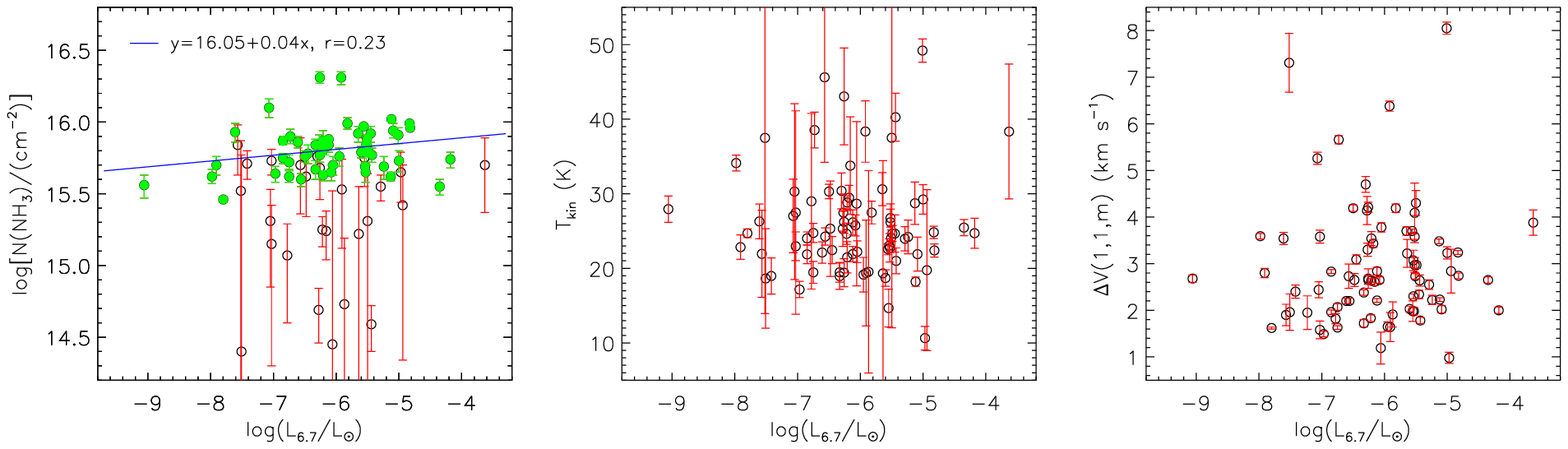}
\caption{Left panel: 6.7 GHz maser luminosity and \nht\ column density. Green filled circles are sources with err$(\log[N(\mathrm{NH}_3)/(\mathrm{cm}^{-2})]) < 0.1$. Blue line is the fitted green circles. Middle panel: 6.7 GHz maser luminosity and \nht\ kinematic temperature. Right panel: 6.7 GHz maser luminosity and \nht$(1,1,m)$ FWHM line width. }\label{fig_maser_nh3}
\end{figure*}

The current sample was also examined for such correlations. The left panel of Figure~\ref{fig_maser_nh3} shows \nht\ column density and maser luminosity, with weak correlation, $r = 0.23$, ignoring sources with large uncertainties. The sharp increase in \nht\ column density at higher maser luminosities, found in \cite{Wu2010}, is not evident in the current sample set. \nht\ kinematic temperature and line width are shown in the middle and right panels, respectively, of Figure~\ref{fig_maser_nh3}, with no significant correlations to maser luminosity, confirming \cite{Pan2012}.

Thus, \nht\ parameters have no evolutionary sequences in the hot phase (Section~\ref{sect:out nh3}), and ammonia properties are independent of the environment around the maser sites.

\begin{figure*}[!t]
\centering
\plotone{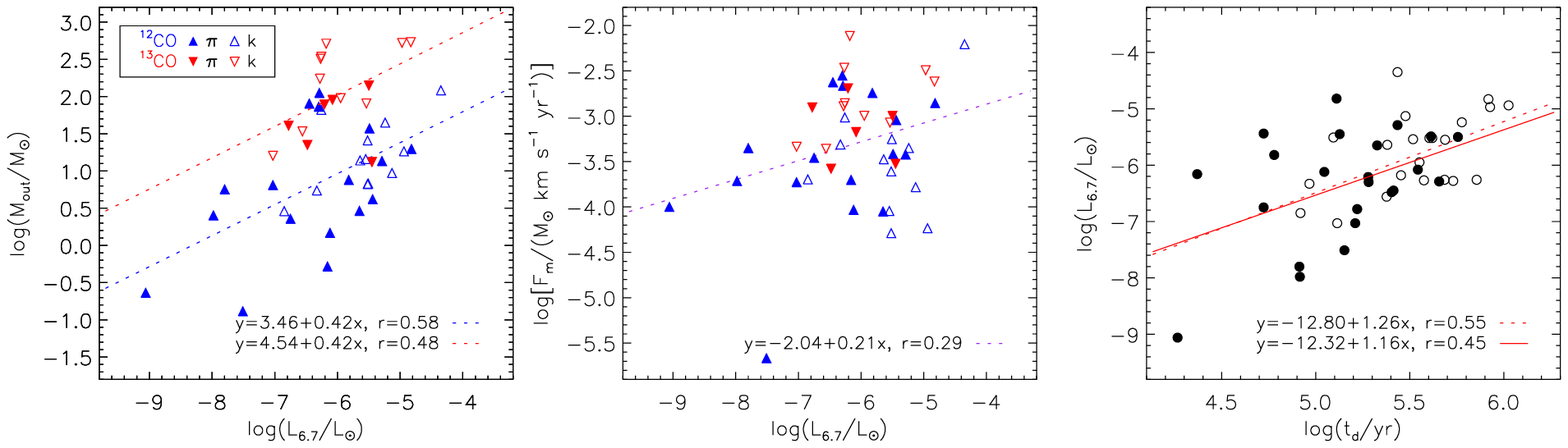}
\caption{Left panel: 6.7 GHz maser luminosity and \coi\ or \coii\ outflow mass. Filled triangles represent sources with accurate parallax distances. Blue and red dashed lines are linear fits for \coi\ and \coii\ outflows, respectively. Middle panel: 6.7 GHz maser luminosity and outflow mechanical force. Purple dashed line is the linear fit. Right panel: 6.7 GHz maser luminosity and outflow dynamical time scale. Solid line is the linear fit for sources with parallax distances (filled circles), dashed line is the fit for all sources.}\label{fig_maser_out}
\end{figure*}

CO outflow properties were investigated for variance with maser luminosity. The left panel in Figure~\ref{fig_maser_out} shows that both \coi\ and \coii\ outflow masses rise with the maser luminosity, $r= 0.58$ and 0.48, respectively, which is consistent with \cite{Vil2015II}. There is also a weak correlation, shown in the middle panel of Figure~\ref{fig_maser_out}, between outflow mechanical force and maser luminosity. As discussed in Section~\ref{sect:out nh3}, outflow driving source may be the pumping mechanism for the 6.7 GHz methanol maser, so a more powerful source driving more massive outflows may pump more maser emission. The right panel shows the evolutionary sequence of the maser luminosity, displaying a weak positive correlation, which is consistent with \cite{Breen2010b}, who stated that maser luminosity increases as the source evolves. However, an evolutionary sequence is not obvious within the sample of \cite{Vil2015II}. Since maser emission disappears before outflow activity stops, the correlation may be due to a distance dependency of both maser luminosity and outflow dynamical time.

\section{Summary and Conclusions}
\label{sect:summary}

Single point observations of \nht~(1,1)~and~(2,2) inversion transitions around 100 6.7 GHz methanol masers were investigated using the Effelsberg 100 m radio telescope, resulting in 82 detections at the 3$\sigma$ level. Forty-five \coi\ or \coii\ outflows from 62 sources were identified from CO $(1-0)$ mapping observations performed on the PMO 13.7 m telescope. Twenty-two of these were new outflow candidates. Temperatures and column densities were calculated from the ammonia lines and outflow parameters estimated without inclination correction. Correlations between source properties, outflow properties, and the maser luminosities were investigated, with the main conclusions:
\begin{enumerate}
\item CO outflows occur around 6.7 GHz maser sites. The detection frequency of outflows was 73\%.
\item Outflow mass and mass outflow rate scale with the IRAS luminosity as well as \coiii\ core mass.
\item Ammonia line properties have no significant evolutionary trends and show no correlation with maser luminosity.
\item Outflow parameters are weakly correlated with maser luminosity, indicating a physical connection of the outflow driving and maser pumping mechanisms.
\end{enumerate}

However, due to the limitations of the telescope resolution, the results and discussions require careful examination; the determination of outflow parameters retains uncertainties; and the correlations may be over dependent on distance. Further studies using higher-level transitions of CO or with larger telescopes or interferometers are expected to improve the results.
%\newpage


%% Included in this acknowledgments section are examples of the
%% AASTeX hypertext markup commands. Use \url without the optional [HREF]
%% argument when you want to print the url directly in the text. Otherwise,
%% use either \url or \anchor, with the HREF as the first argument and the
%% text to be printed in the second.

\acknowledgments
We would thank the referee for their constructive comments which improved the paper. This work was supported by the National Natural Science Foundation of China (grant Nos.: 11133008 and 11233007), the Strategic Priority Research Program of the Chinese Academy of Sciences (grant No.: XDB09010300), and the Key Laboratory for Radio Astronomy.

%%[A16]

\clearpage

\appendix
\section{Calculation of Physical Properties of Ammonia Lines}
\label{sect:A_nh3}
The methods used to calculate physical properties of ammonia lines refer to derivations by \citet{Ho1983}, \cite{Mang1992} and \cite{Pan2012}. All components were assumed to have equal beam filling factors and excitation temperatures, and the radiation process was assumed to occur under conditions of local thermodynamic equilibrium (LTE).

Optical depths are required for further calculations, and these were calculated within the CLASS package when fitting hyperfine components of \nhti\ lines. Opacity can be determined from
\begin{equation}
\label{equ_tau}
\frac{\eqTmb(1,1,m)}{\eqTmb(1,1,s)} = \frac{1 - e^{-\tau (1,1,m)}}{1 - e^{-a \tau (1,1,m)}},
\end{equation}
where \Tmb\ is the main beam temperature; $m$ and $s$ are the main and satellite components, respectively; and $a$ is the relative intensity ratio of the satellite to main component. Here, $a = 0.278$ for $(1,1)$ $F = 1 \rightarrow 2$ and $2 \rightarrow 1$ hyperfine components and $a = 0.222$ for $(1,1)$ $F = 1 \rightarrow 0$ and $0 \rightarrow 1$ hyperfines.

The excitation temperature (\Tex) can be determined from the principle of radiative transfer. Given \Tmb\ and optical depth ($\tau$), \Tex\ was calculated following \citep{Pan2012},
\begin{equation}
\label{equ_Tex}
\eqTmb = \eta_{\mathrm{ff}} [J_{\nu}(\eqTex)-J_{\nu}(T_{\mathrm{bg}})](1-e^{-\tau}),
\end{equation}
where $\eta_{\mathrm{ff}}$ is the beam filling factor, $T_{\mathrm{bg}}$ is the cosmic background temperature, and $J_{\nu}(T)= (h\nu/k) / (e^{h\nu/kT}-1)$. The dense gas is assumed to be uniformly filling the beam, which underestimates \Tex\ due to beam dilution. Therefore, following LTE, taking \Tex\ = \Trot, the beam filling factor was inversely estimated ($\eta_{\mathrm{ff}}$) using Equation~(\ref{equ_Tex}).

Observing \nhti\ and \nhtii\ transitions, the rotational temperature may be calculated to characterize the population distribution between the (1,1) and (2,2) energy levels. \Trot\ was solved from Equation (4) in \cite{Ho1983},
%\begin{multline}
\begin{equation}
\label{equ_Trot}
\eqTrot(2,2;1,1)=-\Delta T\div \ln \biggl\{\frac{-0.282}{\tau(1,1,m)}\ln \biggl[1- \frac{\eqTmb(2,2,m)}{\eqTmb(1,1,m)}(1-e^{-\tau(1,1,m)})\biggr] \biggr\},
\end{equation}
%\end{multline}
where $\Delta T = [E(2,2)-E(1,1)]/k = 41.5~ \mathrm{K}$ is the temperature associated with the energy level difference.

Following \cite{Walm1983}, the kinetic temperature (\Tkin) was determined from \Trot. Assuming a three level system, (1,1), (2,2), and (2,1), and considering the collision rate coefficients from \cite{Danby1988}, \Tkin\ was numerically derived from:
\begin{equation}
\label{equ_Tkin}
1+0.8\exp\left(-\frac{21.5}{\eqTkin}\right)=\exp \left[ 41.5\left( \frac{1}{\eqTrot} - \frac{1}{\eqTkin} \right) \right].
\end{equation}

The ammonia column density was calculated using \cite{Mang1992}. Assuming a Gaussian line profile and the same excitation temperatures for both main and satellite components, the total \nhti\ column density may be obtained by
\begin{equation}
\label{equ_n11}
N(1,1)=6.60 \times 10^{14} \frac{\Delta V(1,1)}{\nu(1,1)} \tau(1,1,m) \eqTex ~\mathrm{cm}^{-2},
\end{equation}
where $\Delta V(1,1)$ is the FWHM velocity, \Tex\ = \Trot\ under LTE, and $\nu(1,1)$ is the frequency of \nhti\ transition. Knowing the partition function, the total column density of \nht\ from $N(1,1)$ may be calculated using
%\begin{multline}
\begin{equation}
\label{equ_nnh3}
N(\mathrm{NH}_3) = N(1,1) \left(1 + \frac{1}{3}e^{23.4/\eqTrot} + \frac{5}{3}e^{-41.5/\eqTrot} + \frac{14}{3}e^{-101.5/\eqTrot} \right).
\end{equation}
%\begin{multline}

Uncertainties were estimated using the Monte Carlo method.

\section{Calculation of Outflow Parameters}
\label{sect:B_co}

The equations derived in \cite{Gard1990} were used to estimate column density. Under LTE and assuming that all levels have the same excitation temperature, the total column density of a linear, rigid rotor molecule from one transition is
\begin{equation}
\label{equ_N}
N = \frac{3k}{8 \pi^3 B \mu^2}\frac{\exp[hBJ(J+1)/k\eqTex]}{J+1}\frac{\eqTex + hB/3k}{1-\exp(-h\nu/k\eqTex)}\int\tau_{\nu} \mathrm{d}v,
\end{equation}
where $B$ and $\mu$ are the rotational constant and permanent dipole moment, respectively; and $J$ is the quantum number of the lower rotational level. For \coi$(J=1-0)$, assuming the high velocity gas to be optical thin and area beam filling factor $f=1$, the total column density in one beam can be given by \citep[see also][]{Snell1984}
\begin{equation}
\label{equ_N12}
N(^{12}\mathrm{CO})=4.2 \times 10^{13} \frac{\eqTex}{\exp(-5.5/\eqTex)}\int \eqTmb \mathrm{d}v,
\end{equation}
where the integrated range is the wing range. The assumption of optically thin high velocity gas may underestimate the column density. For \coii$(J=1-0)$, taking the same assumptions, the \coii\ column density is
\begin{equation}
\label{equ_N13}
N(^{13}\mathrm{CO})=4.6 \times 10^{13} \frac{\eqTex}{\exp(-5.3/\eqTex)}\int \eqTmb \mathrm{d}v.
\end{equation}
The excitation temperature, \Tex\, was assumed to be 30 K here for high-mass sources \citep{Sheph1996a,Beu2002,Wu+2004,Xu2006,Wu2010}.

The column density of H$_2$ gas is required to estimate outflow masses. The conversion factor [H$_2$]/[\coi] $=1 \times 10^4$ was employed for \coi\ lines, whereas [H$_2$]/[\coii] $=5 \times 10^5$ was used for \coii\ lines \citep{Dick1978}.

Given the area of the outflow lobe and the column density, the mass of each lobe was calculated using
\begin{equation}
\label{equ_mbr}
M_{\eqbr} = (N_{\eqbr} \times A_{\eqbr}) m(\mathrm{H}_2),
\end{equation}
where $A_{\eqbr}$ is the blue or red lobe area, and $m$(H$_2$) is the mass of a hydrogen molecule. The total outflow mass was then obtained by adding the masses of each lobe, $M_{\mathrm{out}} = M_{\mathrm{b}} + M_{\mathrm{r}}$.

The outflow velocity relative to the central cloud must first be obtained to calculate outflow momentum and energy. In contrast to \cite{Beu2002}, who measured the outflow velocity from wing extremes, lobe velocities were calculated in each spatial pixel by temperature weighted averaging the velocities in the wing channels,
\begin{equation}
\label{equ_vbr}
\langle\Delta v_{\eqbr}\rangle = \frac{\displaystyle\sum_{i} (v_i-v_{\mathrm{peak}})~ T_i ~\Delta v_{\mathrm{res}}}{\displaystyle\sum_i ~T_i ~\Delta v_{\mathrm{res}}},
\end{equation}
where $i$ is the number of velocity channel in either the blue or red wing, and $\Delta v_{\mathrm{res}}$ is the velocity resolution of a channel. Similarly, the square of lobe velocity is
\begin{equation}
\label{equ_v2}
\langle\Delta v^2_{\eqbr}\rangle = \frac{\displaystyle\sum_{i} (v_i-v_{\mathrm{peak}})^2~ T_i ~\Delta v_{\mathrm{res}}}{\displaystyle\sum_i ~T_i ~\Delta v_{\mathrm{res}}}.
\end{equation}
Outflow momentum and energy were calculated by summing over all the pixels in the lobe area,
\begin{equation}
\label{equ_p}
p_{\mathrm{out}}=\sum_{A_{\mathrm{b}}} M_{\mathrm{b}} \langle \Delta v_{\mathrm{b}} \rangle + \sum_{A_{\mathrm{r}}} M_{\mathrm{r}} \langle \Delta v_{\mathrm{r}} \rangle,
\end{equation}
\begin{equation}
\label{equ_E}
E_{\mathrm{out}}=\frac{1}{2}\sum_{A_{\mathrm{b}}} M_{\mathrm{b}} \langle \Delta v^2_{\mathrm{b}} \rangle + \frac{1}{2}\sum_{A_{\mathrm{r}}} M_{\mathrm{r}} \langle \Delta v^2_{\mathrm{r}} \rangle.
\end{equation}

The dynamical time scale, $t_{\mathrm{d}}$, was calculated by dividing the length of a lobe by the lobe velocity. For a bipolar outflow, $l_{\mathrm{max}} = \max(l_{\mathrm{b}},l_{\mathrm{r}})$ was chosen as the lobe length from the center core, and
\begin{equation}
\label{equ_t}
t_{\mathrm{d}} = \frac{2 l_{\mathrm{max}}}{\Delta v_{\mathrm{b}}+\Delta v_{\mathrm{r}}},
\end{equation}
where $\Delta v_{\eqbr}$ is the averaged lobe velocity in the lobe area. For outflows with only one identified lobe, $t_{\mathrm{d}} = l_{\eqbr}/\Delta v_{\eqbr}$. The mass loss rate, mechanical force, and mechanical luminosity of the molecular outflow were subsequently calculated using
\begin{equation}
\label{equ_m/t}
\dot{M}_{\mathrm{out}} = M_{\mathrm{out}}/t_{\mathrm{d}},
\end{equation}
\begin{equation}
\label{equ_Fm}
F_{\mathrm{m}}=p_{\mathrm{out}}/t_{\mathrm{d}},
\end{equation}
\begin{equation}
\label{equ_Lm}
L_{\mathrm{m}}=E_{\mathrm{out}}/t_{\mathrm{d}}.
\end{equation}

Cloud core mass was estimated using \coiii\ data. Assuming LTE and optically thin emission, the beam averaged column density of \coiii\ \citep[see also][]{Gard1991} is
\begin{equation}
\label{equ_N18}
N(\mathrm{C^{18}O})=4.8 \times 10^{13} \frac{\eqTex}{\exp(-5.3/\eqTex)}\int \eqTmb \mathrm{d}v.
\end{equation}
If the conversion factor [H$_2$]/[\coiii] $=5 \times 10^6$, then the mass of the cloud core is
\begin{equation}
\label{equ_M18}
M_{\mathrm{core}}(\mathrm{C^{18}O})= N_{\mathrm{H}_2}(\mathrm{C^{18}O}) \times A_{\mathrm{core}} \times m(\mathrm{H}_2),
\end{equation}
where $A_{\mathrm{core}}$ is the area of the cloud core.

%% Tables may also be prepared as separate files. See the accompanying
%% sample file table.tex for an example of an external table file.
%% To include an external file in your main document, use the \input
%% command. Uncomment the line below to include table.tex in this
%% sample file. (Note that you will need to comment out the \documentclass,
%% \begin{document}, and \end{document} commands from table.tex if you want
%% to include it in this document.)


%% The following command ends your manuscript. LaTeX will ignore any text

%if emulateapj then
%\LongTables[A17]

\clearpage
\begin{deluxetable}{lccccccccc}
\tabletypesize{\small}
\tablecolumns{8}
\centering
\tablecaption{Basic Properties of the 6.7 GHz Masers Sample\label{tab1}}
\tablewidth{0pt}
\tabcolsep = 6.5 pt
\tablehead{
\colhead{Source Name} & \colhead{R.A.(J2000)} & \colhead{Decl.(J2000)} & \colhead{$S_{\mathrm{peak}}$} &
\colhead{$V_{\mathrm{peak}}$} & \colhead{Distance} & \colhead{Luminosity} & \colhead{Refs}\\
\colhead{} & \colhead{} & \colhead{} & \colhead{(Jy)} & \colhead{(\kms)} & \colhead{(kpc)} & \colhead{(L$_{\odot}$)} &\\
\colhead{(1)} & \colhead{(2)} & \colhead{(3)} & \colhead{(4)} & \colhead{(5)} & \colhead{(6)} & \colhead{(7)} & \colhead{(8)}
}
\startdata
%%SOURCE          EXACT POSITION                      Speak    &     Vpeak    & Dk(Reid)   D|¦°  &    L(6.7GHz) ref (posref (Dpi)
G9.621$+$0.196   & 18:06:14.66  & $-$20:31:31.6  &  5090.0   &  1       &  $^{\pi}$5.15             &  2.36E-04   &    4        \\
G9.986$-$0.028   & 18:07:50.12  & $-$20:18:56.5  &  67.6     &  42.2    &  $^{\mathrm{F}}$11.92     &  1.67E-05   &    1        \\
G12.025$-$0.031  & 18:12:01.86  & $-$18:31:55.7  &  96.3     &  108.3   &  $^{\pi}$9.43             &  1.49E-05   &    1        \\
G12.181$-$0.123  & 18:12:41.00  & $-$18:26:21.9  &  1.9      &  29.7    &  $^{\mathrm{X}}$2.83      &  2.68E-08   &    2        \\
G12.199$-$0.033  & 18:12:23.44  & $-$18:22:50.9  &  12.5     &  49.3    &  $^{\mathrm{F}}$11.77     &  3.02E-06   &    2        \\
G12.202$-$0.120  & 18:12:42.93  & $-$18:25:11.8  &  0.8      &  26.4    &  $^{\mathrm{X}}$2.96      &  1.22E-08   &    2        \\
G12.203$-$0.107  & 18:12:40.24  & $-$18:24:47.5  &  2.4      &  20.5    &  $^{\mathrm{X}}$2.67      &  3.02E-08   &    2        \\
G12.625$-$0.017  & 18:13:11.30  & $-$17:59:57.6  &  25.5     &  21.6    &  $^{\mathrm{N}}$2.35      &  2.46E-07   &    2        \\
G12.681$-$0.182  & 18:13:54.75  & $-$18:01:46.6  &  351.0    &  57.5    &  $^{\pi}$2.40             &  3.54E-06   &    2        \\
G12.889$+$0.489  & 18:11:51.40  & $-$17:31:29.6  &  68.9     &  39.3    &  $^{\pi}$2.50             &  7.51E-07   &    2        \\
G12.909$-$0.260  & 18:14:39.53  & $-$17:52:00.0  &  269.1    &  39.9    &  $^{\pi}$2.53             &  2.99E-06   &    2        \\
G13.657$-$0.599  & 18:17:24.26  & $-$17:22:12.5  &  32.2     &  51.2    &  $^{\mathrm{F}}$12.03     &  8.13E-06   &    2        \\
G14.101$+$0.087  & 18:15:45.81  & $-$06:39:09.4  &  146.0    &  15.2    &  $^{\mathrm{N}}$5.40      &  7.43E-06   &    3        \\
G14.332$-$0.639  & 18:18:53.37  & $-$16:47:39.5  &  0.4      &  21      &  $^{\pi}$1.12             &  8.75E-10   &    4        \\
G14.604$+$0.017  & 18:17:01.14  & $-$16:14:38.0  &  2.3      &  24.6    &  $^{\mathrm{N}}$2.47      &  2.48E-08   &    2        \\
G15.034$-$0.677  & 18:20:24.79  & $-$16:11:35.5  &  39.0     &  21      &  $^{\pi}$1.98             &  2.67E-07   &    4        \\
G16.585$-$0.051  & 18:21:09.13  & $-$14:31:48.5  &  36.7     &  62.1    &  $^{\pi}$3.58             &  8.23E-07   &    2        \\
G16.864$-$2.159  & 18:29:24.42  & $-$15:16:04.5  &  28.9     &  15      &  $^{\mathrm{X}}$1.67      &  1.41E-07   &    2        \\
G17.638$+$0.157  & 18:22:26.30  & $-$13:30:12.1  &  24.8     &  20.8    &  $^{\mathrm{N}}$1.96      &  1.66E-07   &    2        \\
G18.461$-$0.004  & 18:24:36.35  & $-$12:51:08.0  &  23.0     &  49      &  $^{\mathrm{N}}$3.67      &  5.40E-07   &    4        \\
G19.365$-$0.030  & 18:26:25.79  & $-$12:03:52.0  &  33.8     &  25.3    &  $^{\mathrm{N}}$2.15      &  2.73E-07   &    2        \\
G19.472$+$0.170  & 18:25:54.72  & $-$11:52:33.0  &  18.0     &  21.7    &  $^{\mathrm{N}}$1.65      &  8.53E-08   &    2        \\
G19.486$+$0.151  & 18:26:00.39  & $-$11:52:22.6  &  6.0      &  20.6    &  $^{\mathrm{N}}$1.90      &  3.80E-08   &    2        \\
G19.496$+$0.115  & 18:26:09.16  & $-$11:52:51.7  &  7.6      &  121.2   &  $^{\mathrm{F}}$9.63      &  1.22E-06   &    2        \\
G19.701$-$0.267  & 18:27:55.52  & $-$11:52:40.3  &  10.7     &  43.9    &  $^{\mathrm{F}}$12.36     &  2.86E-06   &    2        \\
G20.081$-$0.135  & 18:28:10.32  & $-$11:28:47.0  &  2.0      &  44      &  $^{\mathrm{X}}$12.33     &  5.30E-07   &    4        \\
G20.237$+$0.065  & 18:27:44.56  & $-$11:14:54.2  &  77.0     &  71.8    &  $^{\mathrm{N}}$4.32      &  2.51E-06   &    3        \\
G20.239$+$0.065  & 18:27:44.95  & $-$11:14:48.9  &  5.5      &  70.6    &  $^{\mathrm{N}}$4.31      &  1.78E-07   &    3        \\
G21.880$+$0.014  & 18:31:01.75  & $-$09:49:01.0  &  15.0     &  21.0    &  $^{\mathrm{F}}$13.49     &  4.76E-06   &    1        \\
G22.335$-$0.155  & 18:32:29.40  & $-$09:29:30.1  &  43.0     &  35.7    &  $^{\mathrm{N}}$2.55      &  4.88E-07   &    1        \\
G22.356$+$0.066  & 18:31:44.13  & $-$09:22:12.5  &  12.0     &  79.7    &  $^{\mathrm{N}}$4.72      &  4.66E-07   &    4        \\
G23.010$-$0.411  & 18:34:40.29  & $-$09:00:38.1  &  415.4    &  74.8    &  $^{\pi}$4.59             &  1.52E-05   &    1        \\
G23.207$-$0.378  & 18:34:55.20  & $-$08:49:14.2  &  38.2     &  81.7    &  $^{\mathrm{F}}$10.73     &  7.67E-06   &    1        \\
G23.257$-$0.241  & 18:34:31.26  & $-$08:42:47.0  &  4.4      &  64.1    &  $^{\mathrm{X}}$3.75      &  1.08E-07   &    4        \\
G23.437$-$0.184  & 18:34:39.25  & $-$08:31:38.5  &  45.0     &  103     &  $^{\pi}$5.90             &  2.73E-06   &    3        \\
G23.440$-$0.182  & 18:34:39.18  & $-$08:31:24.3  &  25.0     &  96.6    &  $^{\pi}$5.88             &  1.51E-06   &    3        \\
G23.484$+$0.097  & 18:33:44.05  & $-$08:21:20.5  &  12.0     &  87.4    &  $^{\mathrm{X}}$4.72      &  4.66E-07   &    4        \\
G23.707$-$0.198  & 18:35:12.37  & $-$08:17:39.5  &  9.2      &  79      &  $^{\pi}$6.21             &  6.19E-07   &    4        \\
G24.148$-$0.009  & 18:35:20.94  & $-$07:48:55.6  &  26.8     &  17.7    &  $^{\mathrm{X}}$1.32      &  8.14E-08   &    1        \\
G24.329$+$0.144  & 18:35:08.14  & $-$07:35:04.0  &  5.0      &  110.2   &  $^{\mathrm{F}}$9.31      &  7.56E-07   &    1        \\
G24.493$-$0.039  & 18:36:05.83  & $-$07:31:20.6  &  12.0     &  115     &  $^{\mathrm{X}}$5.65      &  6.68E-07   &    2        \\
G24.790$+$0.083  & 18:36:12.57  & $-$07:12:11.5  &  97.0     &  113     &  $^{\mathrm{F}}$9.38      &  1.49E-05   &    4        \\
G25.411$+$0.105  & 18:37:16.92  & $-$06:38:28.0  &  24.9     &  96      &  $^{\mathrm{X}}$5.07      &  1.12E-06   &    4        \\
G25.650$+$1.050  & 18:34:20.91  & $-$05:59:40.5  &  178.0    &  41.9    &  $^{\mathrm{F}}$12.03     &  4.49E-05   &    4        \\
G25.710$+$0.044  & 18:38:03.15  & $-$06:24:15.0  &  364.0    &  92.8    &  $^{\pi}$10.20            &  6.61E-05   &    4        \\
G25.826$-$0.178  & 18:39:03.63  & $-$06:24:09.5  &  70.0     &  91.2    &  $^{\mathrm{N}}$5.01      &  3.06E-06   &    4        \\
G26.528$-$0.266  & 18:40:40.23  & $-$05:49:07.5  &  9.0      &  104.5   &  $^{\mathrm{F}}$9.29      &  1.35E-06   &    4        \\
G26.602$-$0.220  & 18:40:38.55  & $-$05:43:56.0  &  19.0     &  109     &  $^{\mathrm{F}}$9.19      &  2.80E-06   &    4        \\
G27.220$+$0.261  & 18:40:03.72  & $-$04:57:45.6  &  6.2      &  9.3     &  $^{\mathrm{F}}$13.82     &  2.07E-06   &    1        \\
G27.223$+$0.137  & 18:40:30.43  & $-$05:00:59.0  &  22.0     &  118     &  $^{\mathrm{F}}$8.84      &  3.00E-06   &    4        \\
G27.286$+$0.151  & 18:40:34.48  & $-$04:57:13.5  &  21.0     &  34.6    &  $^{\mathrm{F}}$12.47     &  5.70E-06   &    4        \\
G27.369$-$0.164  & 18:41:50.98  & $-$05:01:28.0  &  29.0     &  99.4    &  $^{\pi}$8.00             &  3.24E-06   &    4        \\
G28.146$-$0.005  & 18:42:42.59  & $-$04:15:36.5  &  61.0     &  101.2   &  $^{\mathrm{N}}$5.25      &  2.93E-06   &    3        \\
G28.201$-$0.049  & 18:42:58.08  & $-$04:13:56.2  &  3.5      &  98.9    &  $^{\mathrm{F}}$9.45      &  5.45E-07   &    3        \\
G28.305$-$0.387  & 18:44:21.99  & $-$04:17:38.5  &  62.0     &  80.7    &  $^{\mathrm{F}}$10.08     &  1.10E-05   &    1        \\
G28.810$+$0.360  & 18:42:37.49  & $-$03:30:12.5  &  6.0      &  91.3    &  $^{\mathrm{F}}$9.55      &  9.54E-07   &    1        \\
G28.829$+$0.488  & 18:42:12.43  & $-$03:25:39.5  &  65.0     &  83.3    &  $^{\mathrm{F}}$9.74      &  1.08E-05   &    4        \\
G28.832$-$0.253  & 18:44:51.09  & $-$03:45:48.0  &  73.0     &  86      &  $^{\mathrm{N}}$4.74      &  2.86E-06   &    4        \\
G29.313$-$0.165  & 18:45:24.97  & $-$03:17:44.5  &  4.0      &  45.5    &  $^{\mathrm{F}}$11.52     &  9.26E-07   &    1        \\
G29.863$-$0.044  & 18:45:59.57  & $-$02:45:04.4  &  76.5     &  101.4   &  $^{\pi}$6.21             &  5.15E-06   &    1        \\
%G29.918$-$0.035  & 18:46:03.69  & $-$02:41:52.5  &  0.0      &  96.8    &  $^{\mathrm{X}}$5.24      &  \nodata    &    1        \\
%G29.923$+$0.059  & 18:45:44.18  & $-$02:39:03.5  &  0.0      &  99.0    &  $^{\mathrm{X}}$5.35      &  \nodata    &    1        \\
G29.956$-$0.016  & 18:46:03.74  & $-$02:39:21.4  &  206.0    &  96      &  $^{\pi}$5.26             &  9.95E-06   &    4        \\
%G30.009$-$0.017  & 18:46:09.85  & $-$02:36:30.5  &  0.0      &  99.0    &  $^{\mathrm{X}}$5.36      &  \nodata    &    1        \\
G30.199$-$0.169  & 18:47:03.07  & $-$02:30:33.6  &  16.0     &  108.6   &  $^{\mathrm{N}}$5.60      &  8.75E-07   &    1        \\
G30.225$-$0.180  & 18:47:08.30  & $-$02:29:27.1  &  10.8     &  113.5   &  $^{\mathrm{X}}$5.64      &  5.99E-07   &    1        \\
G30.591$-$0.042  & 18:47:18.89  & $-$02:06:07.0  &  7.5      &  43      &  $^{\mathrm{N}}$2.60      &  8.84E-08   &    4        \\
G30.761$-$0.052  & 18:47:39.73  & $-$01:57:22.0  &  68.0     &  92.0    &  $^{\mathrm{X}}$5.04      &  3.01E-06   &    1        \\
G30.781$+$0.231  & 18:46:41.52  & $-$01:48:32.0  &  19.0     &  48.9    &  $^{\mathrm{N}}$2.96      &  2.90E-07   &    1        \\
G30.790$+$0.205  & 18:46:48.09  & $-$01:48:46.0  &  23.0     &  86      &  $^{\mathrm{F}}$9.66      &  3.74E-06   &    4        \\
G30.898$+$0.161  & 18:47:09.13  & $-$01:44:10.5  &  2.4      &  102     &  $^{\mathrm{N}}$5.82      &  1.42E-07   &    4        \\
G31.060$+$0.094  & 18:47:41.34  & $-$01:37:21.5  &  6.5      &  16.1    &  $^{\mathrm{X}}$1.08      &  1.32E-08   &    1        \\
G31.282$+$0.062  & 18:48:12.39  & $-$01:26:22.6  &  71.0     &  110     &  $^{\pi}$4.27             &  2.26E-06   &    4        \\
G31.412$+$0.307  & 18:47:34.31  & $-$01:12:47.0  &  11.0     &  104     &  $^{\mathrm{N}}$5.35      &  5.49E-07   &    4        \\
G35.197$-$0.743  & 18:58:13.05  & $+$01:40:35.7  &  72.8     &  28.5    &  $^{\pi}$2.19             &  6.11E-07   &    1        \\
G40.425$+$0.700  & 19:02:39.62  & $+$06:59:10.5  &  15.0     &  15.7    &  $^{\mathrm{F}}$11.43     &  3.42E-06   &    1        \\
G40.623$-$0.138  & 19:06:01.63  & $+$06:46:36.5  &  12.5     &  31.1    &  $^{\mathrm{N}}$2.06      &  9.25E-08   &    3        \\
G41.348$-$0.136  & 19:07:21.87  & $+$07:25:17.3  &  51.0     &  14      &  $^{\mathrm{F}}$11.41     &  1.16E-05   &    4        \\
G43.149$+$0.013  & 19:10:11.06  & $+$09:05:20.0  &  18.0     &  13.5    &  $^{\pi}$11.11            &  3.88E-06   &    1        \\
G43.165$+$0.013  & 19:10:12.89  & $+$09:06:11.9  &  26.0     &  9.3     &  $^{\pi}$11.11            &  5.60E-06   &    1        \\
G43.166$-$0.002  & 19:10:16.72  & $+$09:05:50.6  &  2.4      &  -1.1    &  $^{\pi}$11.11            &  5.17E-07   &    3        \\
G43.171$+$0.004  & 19:10:15.36  & $+$09:06:15.2  &  10.0     &  20.2    &  $^{\pi}$11.11            &  2.15E-06   &    1        \\
G43.796$-$0.127  & 19:11:53.97  & $+$09:35:53.5  &  50.0     &  39.5    &  $^{\pi}$6.02             &  3.16E-06   &    3        \\
G43.890$-$0.784  & 19:14:26.39  & $+$09:22:36.5  &  3.0      &  52      &  $^{\pi}$8.26             &  3.57E-07   &    1        \\
G45.445$+$0.069  & 19:14:18.31  & $+$11:08:59.4  &  0.8      &  50.0    &  $^{\mathrm{F}}$8.60      &  1.08E-07   &    3        \\
G45.467$+$0.053  & 19:14:24.15  & $+$11:09:43.0  &  4.1      &  56.4    &  $^{\pi}$8.40             &  5.05E-07   &    3        \\
G45.473$+$0.133  & 19:14:07.36  & $+$11:12:15.7  &  6.1      &  65.5    &  $^{\mathrm{F}}$7.17      &  5.47E-07   &    3        \\
G45.493$+$0.126  & 19:14:11.35  & $+$11:13:06.2  &  10.0     &  57.1    &  $^{\mathrm{F}}$7.17      &  8.97E-07   &    3        \\
G49.471$-$0.369  & 19:23:37.60  & $+$14:30:05.4  &  4.1      &  73.2    &  $^{\pi}$5.10             &  1.86E-07   &    3        \\
G49.482$-$0.402  & 19:23:46.19  & $+$14:29:47.0  &  7.0      &  50      &  $^{\pi}$5.10             &  3.18E-07   &    3        \\
G49.489$-$0.369  & 19:23:39.83  & $+$14:31:05.0  &  26.0     &  56.1    &  $^{\pi}$5.13             &  1.19E-06   &    3        \\
G49.490$-$0.388  & 19:23:43.95  & $+$14:30:34.4  &  217.0    &  59.2    &  $^{\pi}$5.10             &  9.84E-06   &    1        \\
G52.663$-$1.092  & 19:32:35.30  & $+$16:57:33.0  &  10.0     &  65.0    &  $^{\mathrm{T}}$5.06      &  4.47E-07   &    1        \\
G69.540$-$0.976  & 20:10:09.05  & $+$31:31:35.1  &  31.0     &  14.7    &  $^{\pi}$2.46             &  3.28E-07   &    1        \\
G78.122$+$3.633  & 20:14:26.04  & $+$41:13:33.4  &  38.0     &  $-$6.1  &  $^{\pi}$1.64             &  1.78E-07   &    4        \\
G79.736$+$0.990  & 20:30:50.67  & $+$41:02:27.6  &  18.2     &  $-$5.2  &  $^{\pi}$1.36             &  5.84E-08   &    1        \\
G81.722$+$0.571  & 20:39:01.06  & $+$42:22:49.2  &  2.7      &  $-$3.03 &  $^{\pi}$1.50             &  1.04E-08   &    1        \\
G81.752$+$0.591  & 20:39:01.99  & $+$42:24:59.3  &  4.0      &  $-$9.07 &  $^{\pi}$1.50             &  1.57E-08   &    1        \\
G85.410$+$0.003  & 20:54:13.71  & $+$44:54:07.9  &  42.0     &  $-$29.5 &  $^{\mathrm{T}}$5.60      &  2.30E-06   &    4        \\
G94.603$-$1.796  & 21:39:58.26  & $+$50:14:21.0  &  4.2      &  $-$40.7 &  $^{\pi}$3.57             &  9.34E-08   &    1        \\
G108.184$+$5.519 & 22:28:51.41  & $+$64:13:41.3  &  29.4     &  $-$11   &  $^{\pi}$0.78             &  3.09E-08   &    1        \\
G109.871$+$2.114 & 22:56:18.10  & $+$62:01:49.5  &  815.0    &  $-$2.5  &  $^{\pi}$0.70             &  6.95E-07   &    4        \\
G111.542$+$0.777 & 23:13:45.36  & $+$61:28:10.6  &  296.0    &  $-$56.2 &  $^{\pi}$2.65             &  3.61E-06   &    4
\enddata
\tablecomments{Column (1): The 6.7 GHz methanol maser sources named after galactic coordinates and their alternative names.
Column (2) \& (3): Positions of the 6.7 GHz masers.
Column (4) \& (5): Peak flux densities and peak velocities of the 6.7 GHz masers.
Column (6): Distances of the 6.7 GHz masers, `$\pi$' for trigonometric parallax distances taken from \cite{Reid2014} and `N' or `F' for near or far \citep[see][]{Green2011,Dun2011,Sch2011} kinematic distances calculated from the model A5 given by \cite{Reid2014}. `T' is for those in the tangent regions. We take the near distance for those sources still with ambiguities, which are labelled with `X'.
Column (7): Luminosities calculated from the peak flux densities assuming a typical line width of 0.25~\kms\ and isotropic emissions.
Column (8): References for positions, flux densities and velocities of the 6.7 GHz masers (1.~\citealt{Pest2005}; 2.~\citealt{Green2010ii}; 3.~\citealt{Cas2009}; 4.~\citealt{Xu2009}).}
%\tablerefs{}
%%electronic edition of the {\it Astrophysical Journal}.  A portion is
%%shown here for guidance regarding its form and content.}
%%\tablenotetext{a}{Sample footnote for table that was generated}
%%with the deluxetable environment}
%%\tablenotetext{b}{Another sample footnote for table~\ref{tbl-1}}
\end{deluxetable}

%\newpage
\begin{deluxetable}{lcrrrrr}
\tabletypesize{\small}
\tablecolumns{7}
%\rotate
\tabcolsep = 8 pt
\tablecaption{Fitting Parameters of Ammonia Lines\label{tab2}}
\tablewidth{0pt}
\tablehead{
\colhead{Source Name} & \colhead{$V_{\mathrm{lsr}}$} & \colhead{\Tmb$(1,1)$}& \colhead{$\Delta V(1,1)$} & \colhead{$\tau(1,1,m)$} &
\colhead{\Tmb$(2,2)$} & \colhead{$\Delta V(2,2)$} \\
\colhead{} & \colhead{(\kms)} & \colhead{(K)}& \colhead{(\kms)} & \colhead{} & \colhead{(K)} & \colhead{(\kms)} \\
\colhead{(1)} & \colhead{(2)} & \colhead{(3)} & \colhead{(4)} & \colhead{(5)} & \colhead{(6)} & \colhead{(7)}
}
\startdata
%%!source        & V11       & T11     err  &   dV1  err   &  Tau   err    & T22    err  &    dV2   err    \\
G9.621$+$0.196   &    4.1    &   2.39(0.19) &   3.88(0.27) &   0.90(0.40)  &  1.85(0.16) &    4.06(0.74)   \\
G12.181$-$0.123  &    26.6   &   0.61(0.07) &   1.90(0.23) &   3.22(1.01)  &  0.40(0.07) &    2.17(1.02)   \\
G12.199$-$0.033  &    50.9   &   1.17(0.06) &   2.97(0.16) &   1.74(0.29)  &  0.73(0.05) &    2.61(0.25)   \\
G12.202$-$0.120  &    28.3   &   2.03(0.07) &   2.80(0.10) &   1.59(0.21)  &  1.04(0.07) &    3.18(0.46)   \\
G12.203$-$0.107  &    24.7   &   0.59(0.09) &   7.31(0.63) &   0.32(0.39)  &  0.42(0.05) &    7.13(0.80)   \\
G12.625$-$0.017  &    21.6   &   3.83(0.10) &   2.20(0.06) &   2.89(0.22)  &  2.40(0.11) &    2.61(0.22)   \\
G12.681$-$0.182  &    55.7   &   2.55(0.10) &   2.34(0.09) &   3.07(0.33)  &  1.83(0.11) &    1.95(0.19)   \\
G12.889$+$0.489  &    32.7   &   2.90(0.09) &   2.84(0.08) &   2.04(0.19)  &  1.90(0.09) &    3.30(0.30)   \\
G12.909$-$0.260  &    36.7   &   4.62(0.10) &   3.58(0.06) &   1.62(0.12)  &  2.88(0.11) &    3.30(0.22)   \\
G13.657$-$0.599  &    47.4   &   2.77(0.11) &   2.02(0.08) &   3.86(0.39)  &  1.95(0.12) &    1.99(0.21)   \\
G14.101$+$0.087  &    9.3    &   1.15(0.07) &   3.48(0.05) &   0.98(0.01)  &  0.69(0.06) &    2.96(0.34)   \\
G14.332$-$0.639  &    22.2   &   3.28(0.10) &   2.68(0.08) &   1.11(0.18)  &  1.94(0.10) &    3.06(0.17)   \\
G14.604$+$0.017  &    25.7   &   1.81(0.09) &   3.54(0.13) &   2.04(0.28)  &  1.19(0.07) &    4.12(0.59)   \\
G15.034$-$0.677  &    19.4   &   0.90(0.08) &   2.73(0.26) &   1.15(0.51)  &  0.80(0.09) &    2.07(0.27)   \\
G16.585$-$0.051  &    59.8   &   2.38(0.07) &   2.65(0.07) &   1.44(0.18)  &  1.38(0.06) &    2.59(0.20)   \\
G16.864$-$2.159  &    18.1   &   3.83(0.07) &   2.83(0.04) &   2.25(0.12)  &  2.37(0.07) &    3.22(0.08)   \\
G17.638$+$0.157  &    22.3   &   1.70(0.09) &   1.82(0.11) &   0.52(0.30)  &  0.93(0.08) &    1.90(0.21)   \\
G18.461$-$0.004  &    52.3   &   1.40(0.06) &   3.30(0.14) &   1.74(0.28)  &  0.92(0.06) &    3.46(0.54)   \\
G19.365$-$0.030  &    26.6   &   3.06(0.07) &   2.20(0.06) &   1.55(0.15)  &  1.68(0.08) &    2.57(0.12)   \\
G19.472$+$0.170  &    19.7   &   1.66(0.23) &   5.26(0.13) &   1.97(0.18)  &  1.11(0.05) &    5.75(0.58)   \\
G19.486$+$0.151  &    23.0   &   0.94(0.05) &   2.40(0.14) &   1.93(0.40)  &  0.42(0.06) &    2.02(0.37)   \\
G19.496$+$0.115  &    120.1  &   0.65(0.09) &   1.64(0.31) &   1.88(1.00)  &  0.30(0.09) &    1.10(0.78)   \\
G19.701$-$0.267  &    42.6   &   1.32(0.08) &   3.07(0.22) &   0.10(5.47)  &  0.59(0.06) &    3.39(0.35)   \\
G20.081$-$0.135  &    43.0   &   0.93(0.07) &   4.14(0.30) &   0.10(0.07)  &  0.38(0.05) &    4.42(0.04)   \\
G20.237$+$0.065  &    70.7   &   2.09(0.06) &   2.03(0.06) &   2.76(0.24)  &  1.08(0.06) &    2.02(0.18)   \\
G20.239$+$0.065  &    70.5   &   1.84(0.06) &   2.07(0.07) &   2.31(0.24)  &  0.93(0.05) &    2.92(0.18)   \\
G22.356$+$0.066  &    84.3   &   2.33(0.10) &   1.72(0.08) &   2.46(0.33)  &  1.17(0.10) &    1.94(0.26)   \\
G23.010$-$0.411  &    77.1   &   5.33(0.09) &   2.74(0.06) &   2.91(0.14)  &  3.40(0.10) &    2.95(0.23)   \\
G23.207$-$0.378  &    77.8   &   4.66(0.08) &   2.23(0.04) &   4.24(0.20)  &  2.93(0.10) &    2.91(0.19)   \\
G23.257$-$0.241  &    61.0   &   2.61(0.11) &   1.49(0.06) &   2.66(0.31)  &  1.15(0.09) &    1.91(0.19)   \\
G23.437$-$0.184  &    100.9  &   3.48(0.04) &   3.70(0.04) &   2.20(0.09)  &  1.98(0.07) &    4.47(0.18)   \\
G23.440$-$0.182  &    100.9  &   2.85(0.08) &   4.19(0.09) &   1.93(0.15)  &  1.92(0.08) &    5.03(0.60)   \\
G23.484$+$0.097  &    85.2   &   2.31(0.06) &   2.38(0.06) &   2.59(0.21)  &  1.24(0.07) &    2.85(0.26)   \\
G23.707$-$0.198  &    69.3   &   1.08(0.05) &   3.54(0.15) &   1.87(0.32)  &  0.54(0.05) &    3.03(0.52)   \\
G24.329$+$0.144  &    113.6  &   6.10(0.11) &   2.21(0.04) &   3.07(0.16)  &  3.88(0.12) &    2.58(0.16)   \\
G24.493$-$0.039  &    110.2  &   3.43(0.09) &   3.43(0.08) &   1.53(0.15)  &  2.32(0.09) &    3.04(0.18)   \\
G24.790$+$0.083  &    110.0  &   6.63(0.08) &   3.24(0.04) &   2.59(0.09)  &  4.49(0.12) &    3.36(0.15)   \\
G25.411$+$0.105  &    95.3   &   1.99(0.11) &   1.65(0.10) &   3.18(0.44)  &  1.15(0.11) &    1.89(0.40)   \\
G25.650$+$1.050  &    42.7   &   3.36(0.08) &   2.65(0.07) &   1.14(0.14)  &  1.79(0.07) &    3.56(0.17)   \\
G25.710$+$0.044  &    98.2   &   2.78(0.09) &   2.00(0.07) &   2.33(0.25)  &  1.80(0.11) &    2.28(0.19)   \\
G25.826$-$0.178  &    93.8   &   3.55(0.07) &   2.73(0.06) &   2.42(0.16)  &  2.26(0.08) &    3.72(0.18)   \\
G26.528$-$0.266  &    105.8  &   0.77(0.09) &   1.91(0.27) &   0.25(0.55)  &  0.22(0.05) &    2.03(0.62)   \\
G26.602$-$0.220  &    107.9  &   0.84(0.11) &   1.98(0.22) &   2.56(0.87)  &  0.26(0.05) &    1.69(0.22)   \\
G27.223$+$0.137  &    112.8  &   0.62(0.09) &   4.09(0.64) &   0.10(2.00)  &  0.23(0.09) &    0.89(0.24)   \\
G27.286$+$0.151  &    31.2   &   1.57(0.07) &   2.22(0.10) &   1.92(0.32)  &  0.92(0.07) &    3.08(0.57)   \\
G27.369$-$0.164  &    91.3   &   3.67(0.09) &   2.97(0.07) &   1.83(0.14)  &  2.17(0.08) &    3.92(0.16)   \\
G28.146$-$0.005  &    98.4   &   1.64(0.06) &   1.98(0.08) &   1.97(0.26)  &  0.91(0.07) &    2.14(0.21)   \\
G28.201$-$0.049  &    95.8   &   3.13(0.09) &   2.65(0.09) &   1.88(0.18)  &  2.00(0.09) &    3.50(0.20)   \\
G28.829$+$0.488  &    86.8   &   1.25(0.15) &   0.98(0.12) &   3.53(1.31)  &  0.19(0.03) &    1.24(0.15)   \\
G28.832$-$0.253  &    86.9   &   4.23(0.09) &   2.30(0.05) &   1.86(0.13)  &  2.32(0.09) &    2.90(0.20)   \\
G29.863$-$0.044  &    101.1  &   1.30(0.05) &   2.55(0.10) &   1.19(0.22)  &  0.65(0.04) &    2.40(0.25)   \\
G29.956$-$0.016  &    97.1   &   2.84(0.09) &   3.23(0.13) &   1.35(0.19)  &  1.85(0.09) &    3.96(0.26)   \\
G30.199$-$0.169  &    103.1  &   0.75(0.11) &   1.19(0.34) &   0.19(2.39)  &  0.38(0.08) &    2.27(0.53)   \\
G30.225$-$0.180  &    103.9  &   1.57(0.05) &   1.83(0.07) &   0.84(0.19)  &  0.75(0.04) &    2.27(0.16)   \\
G30.591$-$0.042  &    41.8   &   1.10(0.08) &   2.44(0.17) &   0.66(0.37)  &  0.66(0.04) &    2.90(0.64)   \\
G30.790$+$0.205  &    81.7   &   2.40(0.09) &   1.78(0.07) &   2.94(0.29)  &  1.42(0.08) &    2.15(0.21)   \\
G30.898$+$0.161  &    105.5  &   3.31(0.07) &   1.96(0.05) &   2.54(0.18)  &  1.93(0.08) &    2.43(0.18)   \\
G31.282$+$0.062  &    109.4  &   2.17(0.08) &   3.70(0.09) &   1.76(0.18)  &  1.57(0.06) &    3.88(0.15)   \\
G31.412$+$0.307  &    97.0   &   1.99(0.08) &   4.21(0.11) &   4.35(0.35)  &  1.37(0.08) &    3.29(0.25)   \\
G35.197$-$0.743  &    33.9   &   4.87(0.07) &   2.64(0.04) &   1.29(0.09)  &  3.09(0.07) &    3.21(0.08)   \\
G40.623$-$0.138  &    32.6   &   1.04(0.04) &   3.58(0.14) &   1.32(0.25)  &  0.51(0.04) &    3.83(0.33)   \\
G41.348$-$0.136  &    13.3   &   0.67(0.09) &   2.84(0.47) &   0.84(0.73)  &  0.24(0.10) &    0.51(0.12)   \\
G43.166$-$0.002  &    0.2    &   0.43(0.12) &   1.99(0.57) &   0.10(0.53)  &  0.55(0.10) &    2.24(0.52)   \\
G43.796$-$0.127  &    44.2   &   0.60(0.05) &   4.30(0.27) &   0.33(0.34)  &  0.42(0.03) &    5.26(0.42)   \\
G43.890$-$0.784  &    54.4   &   1.66(0.06) &   3.09(0.10) &   1.70(0.22)  &  0.85(0.06) &    3.25(0.42)   \\
G45.467$+$0.053  &    61.9   &   1.05(0.04) &   4.70(0.17) &   0.96(0.20)  &  0.67(0.04) &    4.08(0.32)   \\
G45.473$+$0.133  &    60.9   &   0.73(0.05) &   2.68(0.21) &   1.16(0.39)  &  0.63(0.04) &    3.01(0.23)   \\
G45.493$+$0.126  &    60.8   &   1.70(0.06) &   3.78(0.10) &   1.18(0.18)  &  0.75(0.05) &    3.94(0.71)   \\
G49.471$-$0.369  &    67.3   &   2.59(0.10) &   5.66(0.09) &   0.99(0.10)  &  2.03(0.04) &    5.96(0.14)   \\
G49.482$-$0.402  &    54.6   &   4.44(0.07) &   4.19(0.07) &   1.08(0.09)  &  2.86(0.07) &    4.84(0.12)   \\
G49.489$-$0.369  &    59.7   &   1.91(0.06) &   6.38(0.11) &   2.25(0.10)  &  1.68(0.06) &    5.92(0.38)   \\
G49.490$-$0.388  &    55.5   &   4.35(0.05) &   8.05(0.13) &   0.61(0.06)  &  3.92(0.06) &    7.35(0.10)   \\
G69.540$-$0.976  &    11.5   &   3.91(0.08) &   2.65(0.14) &   1.35(0.59)  &  2.17(0.07) &    3.40(0.12)   \\
G78.122$+$3.633  &  $-$4.1   &   5.01(0.13) &   1.63(0.04) &   2.18(0.18)  &  3.16(0.12) &    1.47(0.08)   \\
G79.736$+$0.990  &  $-$1.4   &   0.74(0.11) &   1.95(0.36) &   1.17(0.86)  &  1.29(0.49) &    0.79(0.37)   \\
G81.722$+$0.571  &  $-$3.3   &   5.51(0.09) &   3.59(0.06) &   0.87(0.09)  &  3.84(0.08) &    3.58(0.13)   \\
G81.752$+$0.591  &  $-$4.0   &   7.45(0.08) &   1.62(0.02) &   1.54(0.07)  &  4.17(0.09) &    1.88(0.07)   \\
G85.410$+$0.003  &  $-$36.2  &   0.64(0.05) &   3.22(0.30) &   0.46(0.46)  &  0.19(0.05) &    4.04(1.39)   \\
G94.603$-$1.796  &  $-$43.8  &   0.74(0.08) &   1.58(0.19) &   0.73(0.56)  &  0.40(0.08) &    1.55(0.31)   \\
G108.184$+$5.519 &  $-$9.8   &   0.92(0.10) &   1.96(0.39) &   0.12(3.75)  &  0.23(0.07) &    2.90(0.52)   \\
G109.871$+$2.114 &  $-$11.0  &   3.20(0.07) &   2.61(0.08) &   0.50(0.16)  &  2.08(0.08) &    2.45(0.17)   \\
G111.542$+$0.777 &  $-$56.8  &   1.34(0.06) &   2.64(0.12) &   0.10(0.06)  &  0.98(0.06) &    3.01(0.23)
\enddata
\tablecomments{The columns are the fit parameters of \nht~$(1,1)$, $(2,2)$ lines with their uncertainties by the CLASS/GILDAS software.
Column (2) -- (5): LSR velocities, the main-beam temperatures, line widths and opacities for the main component of \nhti\ lines using the `NH3(1,1)' method.
Column (6) \& (7): The main-beam temperatures and line widths for \nhtii\ lines using the `GAUSS' method.}
\end{deluxetable}

%\newpage
\begin{deluxetable}{lccccr}
\tabletypesize{\small}
\tablecolumns{6}
\tabcolsep = 12.5 pt
\tablecaption{Derived Physical Properties of Ammonia Lines\label{tab3}}
\tablewidth{0pt}
\tablehead{
\colhead{Source Name} &  \colhead{\Tex} & \colhead{\Trot} & \colhead{\Tkin} & \colhead{$N$(\nht)} & \colhead{$\eta_\mathrm{ff}$} \\
\colhead{}  & \colhead{(K)} & \colhead{(K)} & \colhead{(K)} & \colhead{($10^{14}$cm$^{-2}$)} &  \colhead{}
}
\startdata
%%!source        &   Tex             Trot            Tkin          [NNH3]/1E14
G9.621$+$0.196   &   6.76(4.82)  &    25.21(3.83) &   38.34(9.04) &  54.82(26.48)  &    0.18  \\
G12.181$-$0.123  &   3.34(0.23)  &    16.91(3.42) &   21.94(5.84) &  71.05(26.00)  &    0.04  \\
G12.199$-$0.033  &   4.13(0.19)  &    19.31(1.32) &   26.18(2.46) &  64.48(11.87)  &    0.09  \\
G12.202$-$0.120  &   5.27(0.25)  &    17.43(0.85) &   22.84(1.64) &  52.42(7.39)   &    0.17  \\
G12.203$-$0.107  &   4.87(11.71) &    24.84(9.09) &   37.49(23.55)&  36.47(41.25)  &    0.10  \\
G12.625$-$0.017  &   6.78(0.24)  &    17.01(0.71) &   22.12(1.43) &  74.19(6.25)   &    0.28  \\
G12.681$-$0.182  &   5.39(0.24)  &    18.47(1.34) &   24.66(2.50) &  87.58(10.68)  &    0.17  \\
G12.889$+$0.489  &   6.06(0.25)  &    19.30(0.92) &   26.16(1.69) &  72.48(7.57)   &    0.20  \\
G12.909$-$0.260  &   8.49(0.36)  &    19.62(0.64) &   26.74(1.20) &  73.48(5.66)   &    0.34  \\
G13.657$-$0.599  &   5.55(0.24)  &    16.88(1.17) &   21.90(2.28) &  90.58(10.16)  &    0.20  \\
G14.101$+$0.087  &   4.56(0.26)  &    20.67(1.46) &   28.73(2.84) &  44.66(2.53)   &    0.10  \\
G14.332$-$0.639  &   7.63(0.62)  &    20.26(0.92) &   27.94(1.76) &  38.35(6.64)   &    0.28  \\
G14.604$+$0.017  &   4.80(0.25)  &    19.37(1.27) &   26.29(2.33) &  90.41(13.27)  &    0.13  \\
G15.034$-$0.677  &   4.04(4.12)  &    28.16(4.18) &   45.61(11.40)&  56.02(27.20)  &    0.05  \\
G16.585$-$0.051  &   5.85(0.27)  &    19.08(0.77) &   25.75(1.41) &  47.29(6.08)   &    0.19  \\
G16.864$-$2.159  &   7.01(0.19)  &    18.11(0.49) &   24.02(0.87) &  76.59(4.40)   &    0.28  \\
G17.638$+$0.157  &   6.93(8.56)  &    20.80(5.88) &   28.99(11.77)&  12.45(7.70)   &    0.23  \\
G18.461$-$0.004  &   4.41(0.22)  &    20.01(1.31) &   27.47(2.49) &  73.61(12.43)  &    0.10  \\
G19.365$-$0.030  &   6.61(0.25)  &    18.25(0.65) &   24.26(1.16) &  41.12(4.12)   &    0.25  \\
G19.472$+$0.170  &   4.65(0.25)  &    19.77(2.56) &   27.02(4.94) & 131.59(17.42)  &    0.11  \\
G19.486$+$0.151  &   3.80(0.18)  &    15.63(1.35) &   18.97(2.42) &  52.20(11.31)  &    0.08  \\
G19.496$+$0.115  &   3.47(4.06)  &    15.90(4.60) &   19.38(7.09) &  34.92(20.88)  &    0.06  \\
G19.701$-$0.267  &       \nodata &    \nodata     &    \nodata    &   \nodata      &  \nodata  \\
G20.081$-$0.135  &  12.53(2.00)  &    18.83(1.27) &   25.30(2.35) &   5.09(1.97)   &    0.61  \\
G20.237$+$0.065  &   4.95(0.14)  &    15.44(0.63) &   18.70(1.11) &  62.93(5.69)   &    0.18  \\
G20.239$+$0.065  &   4.76(0.15)  &    15.96(0.71) &   19.46(1.47) &  54.15(5.92)   &    0.15  \\
G22.356$+$0.066  &   5.27(0.22)  &    15.61(0.95) &   18.95(1.77) &  47.59(6.79)   &    0.20  \\
G23.010$-$0.411  &   8.36(0.32)  &    17.17(0.47) &   22.40(0.88) &  93.50(5.02)   &    0.39  \\
G23.207$-$0.378  &   7.46(0.20)  &    15.10(0.43) &   18.21(0.63) & 105.45(5.35)   &    0.38  \\
G23.257$-$0.241  &   5.52(0.21)  &    14.37(0.74) &   17.16(1.08) &  43.79(5.34)   &    0.24  \\
G23.437$-$0.184  &   6.63(0.14)  &    17.27(0.41) &   22.56(0.73) &  95.36(4.03)   &    0.27  \\
G23.440$-$0.182  &   6.06(0.33)  &    20.01(0.82) &   27.48(1.52) & 103.65(8.85)   &    0.19  \\
G23.484$+$0.097  &   5.22(0.16)  &    15.97(0.62) &   19.47(1.33) &  69.97(6.04)   &    0.19  \\
G23.707$-$0.198  &   3.99(0.18)  &    16.63(1.07) &   21.48(2.10) &  76.19(13.68)  &    0.09  \\
G24.329$+$0.144  &   9.12(0.31)  &    16.88(0.49) &   21.91(1.04) &  78.76(4.49)   &    0.45  \\
G24.493$-$0.039  &   7.11(0.33)  &    21.05(0.85) &   29.47(1.67) &  69.81(7.30)   &    0.24  \\
G24.790$+$0.083  &   9.89(0.24)  &    18.57(0.47) &   24.83(0.83) & 102.35(4.21)   &    0.45  \\
G25.411$+$0.105  &   4.79(0.21)  &    15.75(1.26) &   19.15(2.32) &  59.20(9.00)   &    0.16  \\
G25.650$+$1.050  &   7.67(0.43)  &    18.92(0.60) &   25.46(1.09) &  37.23(4.65)   &    0.31  \\
G25.710$+$0.044  &   5.80(0.23)  &    18.50(1.10) &   24.71(2.02) &  56.84(6.62)   &    0.20  \\
G25.826$-$0.178  &   6.62(0.21)  &    18.14(0.62) &   24.06(1.10) &  79.49(5.63)   &    0.25  \\
G26.528$-$0.266  &   6.20(11.39) &    16.00(8.45) &   19.53(13.57)&   5.44(10.07)  &    0.26  \\
G26.602$-$0.220  &   3.62(1.05)  &    12.59(2.07) &   14.67(2.53) &  56.14(20.59)  &    0.09  \\
G27.223$+$0.137  &      \nodata  & \nodata        &    \nodata    &   \nodata      &   \nodata  \\
G27.286$+$0.151  &   4.55(0.22)  &    18.22(1.23) &   24.21(2.27) &  51.40(9.03)   &    0.12  \\
G27.369$-$0.164  &   7.09(0.28)  &    18.48(0.59) &   24.67(1.06) &  66.19(5.66)   &    0.28  \\
G28.146$-$0.005  &   4.63(0.16)  &    17.39(0.99) &   22.76(1.91) &  46.02(6.32)   &    0.13  \\
G28.201$-$0.049  &   6.41(0.27)  &    19.34(0.82) &   26.23(1.52) &  62.33(6.48)   &    0.22  \\
G28.829$+$0.488  &   4.00(1.82)  &     9.50(2.07) &   10.64(1.59) &  44.01(17.03)  &    0.19  \\
G28.832$-$0.253  &   7.74(0.26)  &    17.55(0.51) &   23.04(0.92) &  50.77(3.75)   &    0.34  \\
G29.863$-$0.044  &   4.58(0.25)  &    18.08(0.89) &   23.96(1.62) &  36.60(6.92)   &    0.12  \\
G29.956$-$0.016  &   6.57(0.42)  &    20.93(1.01) &   29.24(1.97) &  57.78(8.93)   &    0.21  \\
G30.199$-$0.169  &   7.02(6.22)  &    20.63(5.67) &   28.67(11.01)&   3.01(30.59)  &    0.24  \\
G30.225$-$0.180  &   5.50(0.58)  &    18.46(0.79) &   24.63(1.42) &  18.53(4.22)   &    0.18  \\
G30.591$-$0.042  &   5.00(7.30)  &    21.46(5.75) &   30.29(11.79)&  21.81(13.33)  &    0.12  \\
G30.790$+$0.205  &   5.25(0.18)  &    16.34(0.83) &   21.00(1.74) &  60.12(6.48)   &    0.19  \\
G30.898$+$0.161  &   6.31(0.17)  &    16.86(0.59) &   21.87(1.25) &  57.84(4.42)   &    0.25  \\
G31.282$+$0.062  &   5.34(0.24)  &    21.62(1.11) &   30.61(2.22) &  89.03(10.21)  &    0.14  \\
G31.412$+$0.307  &   4.73(0.21)  &    15.93(0.96) &   19.42(1.87) & 207.70(17.85)  &    0.15  \\
G35.197$-$0.743  &   9.44(0.33)  &    20.72(0.47) &   28.83(0.91) &  45.05(3.37)   &    0.37  \\
G40.623$-$0.138  &   4.14(0.22)  &    17.50(1.02) &   22.96(1.92) &  56.00(11.14)  &    0.10  \\
G41.348$-$0.136  &   3.90(7.79)  &    16.15(6.99) &   19.75(10.80)&  27.22(24.38)  &    0.09  \\
G43.166$-$0.002  &     \nodata   & \nodata        &    \nodata    &   \nodata      &  \nodata \\
G43.796$-$0.127  &   4.83(11.04) &    24.85(9.33) &   37.50(25.49)&  22.19(23.00)  &    0.10  \\
G43.890$-$0.784  &   4.75(0.21)  &    17.20(0.93) &   22.44(1.82) &  61.57(8.48)   &    0.14  \\
G45.467$+$0.053  &   4.43(0.33)  &    21.51(1.19) &   30.39(2.40) &  60.93(13.26)  &    0.09  \\
G45.473$+$0.133  &   3.77(1.75)  &    27.17(2.57) &   43.05(6.50) &  53.46(19.24)  &    0.04  \\
G45.493$+$0.126  &   5.18(0.30)  &    17.07(0.74) &   22.22(1.47) &  51.81(7.83)   &    0.17  \\
G49.471$-$0.369  &   6.85(0.49)  &    25.30(1.04) &   38.54(2.40) &  88.48(9.53)   &    0.18  \\
G49.482$-$0.402  &   9.46(0.44)  &    21.46(0.48) &   30.30(0.96) &  61.17(5.52)   &    0.36  \\
G49.489$-$0.369  &   4.85(0.13)  &    25.22(1.78) &   38.35(4.12) & 226.65(20.31)  &    0.09  \\
G49.490$-$0.388  &  12.30(0.84)  &    29.49(0.56) &   49.20(1.56) &  92.43(10.02)  &    0.36  \\
G69.540$-$0.976  &   8.01(5.41)  &    18.84(3.44) &   25.32(6.39) &  44.04(20.07)  &    0.33  \\
G78.122$+$3.633  &   8.38(0.28)  &    18.52(0.71) &   24.74(1.28) &  43.15(3.87)   &    0.36  \\
G79.736$+$0.990  &   3.78(5.69)  & \nodata        &    \nodata    &   \nodata      &  \nodata \\
G81.722$+$0.571  &  12.24(0.74)  &    23.31(0.50) &   34.11(1.06) &  45.35(4.64)   &    0.46  \\
G81.752$+$0.591  &  12.21(0.27)  &    18.50(0.32) &   24.70(0.58) &  30.36(1.56)   &    0.60  \\
G85.410$+$0.003  &   4.43(10.39) &    15.87(9.23) &   19.33(15.09)&  16.89(18.44)  &    0.13  \\
G94.603$-$1.796  &   4.14(7.18)  &    20.01(7.11) &   27.48(13.64)&  14.86(12.06)  &    0.08  \\
G108.184$+$5.519 &  11.12(4.24)  &    15.40(4.47) &   18.64(6.67) &   2.55(56.42)  &    0.66  \\
G109.871$+$2.114 &  10.91(4.52)  &    23.15(2.99) &   33.77(6.52) &  18.84(6.43)   &    0.40  \\
G111.542$+$0.777 &  16.83(2.21)  &    26.03(1.34) &   40.26(3.21) &   4.32(1.39)   &    0.61
\enddata
\tablecomments{The columns present the excitation temperatures, rotational temperatures, kinematic temperatures, column densities and filling factors. Their derivations and calculations refer to Equation~(\ref{equ_Tex}), (\ref{equ_Trot}), (\ref{equ_Tkin}), (\ref{equ_nnh3}) and (\ref{equ_Tex}), respectively.}
\end{deluxetable}

\begin{deluxetable}{lccccccrrrc}
\tabletypesize{\scriptsize}
\tablecolumns{11}
\tabcolsep = 10 pt
\rotate
\centering
\tablecaption{Parameters for Outflow Calculations\label{tab4}}
\tablewidth{0pt}
\tablehead{
\colhead{Source Name}  & \colhead{$v'_{\mathrm{peak}}$} & \colhead{$\Delta V_{\mathrm{b}}$}& \colhead{$\Delta V_{\mathrm{r}}$}
& \colhead{$A_{\mathrm{b}}$} & \colhead{$A_{\mathrm{r}}$} &
 \colhead{$l_{\mathrm{b}}$} & \colhead{$l_{\mathrm{r}}$} & \colhead{$I_{\mathrm{b}}$} & \colhead{$I_{\mathrm{r}}$} & \colhead{New?} \\
\colhead{}  & \colhead{(\kms)} & \colhead{(\kms)} & \colhead{(\kms)} &
 \colhead{(pc$^2$)} & \colhead{(pc$^2$)} & \colhead{(pc)} & \colhead{(pc)} &  \colhead{(K \kms)} & \colhead{(K \kms)} & \\
 \colhead{(1)} & \colhead{(2)} & \colhead{(3)} & \colhead{(4)} & \colhead{(5)} & \colhead{(6)} & \colhead{(7)} & \colhead{(8)} & \colhead{(9)} & \colhead{(10)} & \colhead{(11)} }
\startdata
%Num    l  b line                                 (bv1,bv2)          (rv1,rv2)       area_b  area_r      scale_b     scale_r     I_b         I_r
%                                                                                [pc^2]     [pc^2]       [pc]        [pc]       [K km/s]   [K km/s]
 G12.199$-$0.033$^{ }$    &    51.69  &  \nodata          &  $(54   ,61  )$   & \nodata   &   11.72  & \nodata  &     1.70 & \nodata &     9.11 &    yes  \\
 G12.681$-$0.182$^{*}$    &    56.36  &  \nodata          &  $(58.5 ,65  )$   & \nodata   &    0.61  & \nodata  &     0.42 & \nodata &     4.49 &    yes  \\
 G12.889$+$0.489$^{ }$    &    32.86  &  \nodata          &  $(37   ,47  )$   & \nodata   &    0.53  & \nodata  &     0.79 & \nodata &    11.59 &    yes  \\
 G14.101$+$0.087$^{ }$    &     8.78  &  (0,5)            &  \nodata          &    4.32   & \nodata  &    1.61  &  \nodata &    9.03 &  \nodata &    yes  \\
 G14.332$-$0.639$^{ }$    &    21.86  &  (9,16)           &  \nodata          &    0.08   & \nodata  &    0.15  &  \nodata &   11.85 &  \nodata &    yes  \\
 G16.585$-$0.051$^{*}$    &    59.53  &  (55,57.5)        &  $(61   ,64	 )$   &    1.63   &    1.36  &    0.87  &     0.55 &    5.29 &     3.80 &    no   \\
 G16.864$-$2.159$^{ }$    &    19.33  &  (4,15)           &  $(22   ,32  )$   &    0.53   &    0.35  &    0.26  &     0.47 &   14.48 &    12.25 &    yes  \\
 G17.638$+$0.157$^{*}$    &    22.36  &  (11,19)          &  $(25   ,32  )$   &    0.65   &    0.65  &    0.86  &     0.69 &    6.11 &     5.76 &    no   \\
 G18.461$-$0.004$^{*}$    &    51.53  &  (43,50)          &  $(53.5 ,60  )$   &    1.42   &    3.99  &    0.90  &     1.54 &    7.87 &    12.34 &    yes  \\
 G19.365$-$0.030$^{*}$    &    26.67  &  (18,25)          &  $(29   ,32  )$   &    0.59   &    0.59  &    0.64  &     0.74 &    9.42 &     3.57 &    yes  \\
 G20.081$-$0.135$^{*}$    &    42.67  &  (35,40)          &  $(45   ,50  )$   &    3.22   &    6.43  &    1.21  &     2.18 &    4.30 &     5.19 &    no   \\
 G23.010$-$0.411$^{ }$    &    77.45  &  \nodata          &  $(82   ,96  )$   & \nodata   &    4.46  & \nodata  &     1.21 & \nodata &    18.40 &    no   \\
 G23.440$-$0.182$^{ }$    &   101.28  &  \nodata          &  $(112  ,122 )$   & \nodata   &    2.95  & \nodata  &     0.88 & \nodata &    10.72 &    no   \\
 G23.484$+$0.097$^{ }$    &    85.61  &  (72,80)          &  \nodata          &    1.89   & \nodata  &    0.78  &  \nodata &   12.06 &  \nodata &    no   \\
 G24.493$-$0.039$^{*}$    &   110.36  &  (102,108)        &  $(113  ,120 )$   &    6.08   &    5.40  &    1.17  &     1.25 &   10.57 &     5.43 &    no   \\
 G24.790$+$0.083$^{*}$    &   110.70  &  \nodata          &  $(113  ,119 )$   & \nodata   &   14.89  & \nodata  &     3.14 & \nodata &     5.89 &    no   \\
 G25.411$+$0.105$^{*}$    &    95.86  &  (90,93.5)        &  $(97   ,103 )$   &    1.63   &    2.72  &    0.98  &     1.37 &    2.41 &     6.35 &    no   \\
 G25.650$+$1.050$^{ }$    &    43.09  &  (18,32)          &  $(53   ,67  )$   &   19.74   &   19.74  &    3.85  &     2.62 &   14.02 &    11.53 &    no   \\
 G25.826$-$0.178$^{ }$    &    94.11  &  (78,88)          &  \nodata          &    2.65   & \nodata  &    1.30  &  \nodata &   10.61 &  \nodata &    no   \\
 G26.602$-$0.220$^{ }$    &   108.20  &  (104,106)        &  $(110  ,114 )$   &    7.15   &    7.15  &    1.14  &     1.52 &    4.12 &     4.29 &    yes  \\
 G27.223$+$0.137$^{ }$    &   112.51  &  (107,111)        &  $(115  ,120 )$   &    6.61   &    3.31  &    1.35  &     1.35 &    2.78 &     2.81 &    yes  \\
 G27.286$+$0.151$^{ }$    &    32.36  &  (23,28)          &  \nodata          &   16.45   & \nodata  &    3.65  &  \nodata &   11.30 &  \nodata &    yes  \\
 G27.369$-$0.164$^{ }$    &    92.53  &  (83,88)          &  $(95   ,99  )$   &    6.77   &    9.48  &    1.19  &     1.97 &    6.48 &    11.76 &    yes  \\
 G28.201$-$0.049$^{*}$    &    95.50  &  (90,94)          &  \nodata          &   13.22   & \nodata  &    2.12  &  \nodata &    5.38 &  \nodata &    no   \\
 G28.829$+$0.488$^{*}$    &    86.00  &  (76,83.5)        &  $(88.5 ,96  )$   &   14.05   &   18.06  &    4.06  &     4.57 &    5.70 &     3.25 &    yes  \\
 G28.832$-$0.253$^{*}$    &    87.53  &  (81,85)          &  \nodata          &    3.33   & \nodata  &    1.17  &  \nodata &    4.39 &  \nodata &    no   \\
 G29.863$-$0.044$^{ }$    &   100.42  &  \nodata          &  $(105  ,120 )$   & \nodata   &    1.63  & \nodata  &     2.10 & \nodata &    34.67 &    no   \\
 G31.282$+$0.062$^{ }$    &   108.17  &  \nodata          &  $(113.5,118 )$   & \nodata   &    2.31  & \nodata  &     1.41 & \nodata &     5.22 &    no   \\
 G35.197$-$0.743$^{*}$    &    34.03  &  (24,31)          &  $(38   ,45  )$   &    1.42   &    1.52  &    0.94  &     1.01 &    5.95 &     2.88 &    no   \\
 G40.623$-$0.138$^{*}$    &    32.86  &  (24,30)          &  $(35   ,40  )$   &    0.54   &    0.54  &    0.35  &     0.50 &    3.01 &     3.84 &    no   \\
 G41.348$-$0.136$^{ }$    &    12.50  &  (6,10)           &  $(15   ,18  )$   &    8.26   &   11.02  &    3.66  &     1.57 &    3.92 &     4.00 &    yes  \\
 G43.166$-$0.002$^{ }$    &     2.61  &  ($-$15,$-$3)     &  \nodata          &   23.50   & \nodata  &    4.04  &  \nodata &   19.80 &  \nodata &    no   \\
 G43.796$-$0.127$^{*}$    &    43.34  &  (37,41)          &  $(47   ,52  )$   &    4.60   &    4.60  &    2.10  &     2.35 &    3.31 &     3.81 &    yes  \\
 G43.890$-$0.784$^{ }$    &    55.00  &  (40,50)          &  $(58   ,68.5)$   &    8.66   &    8.66  &    2.02  &     2.02 &   19.41 &    19.18 &    yes  \\
 G45.467$+$0.053$^{ }$    &    60.28  &  (41,51)          &  $(63.5 ,72  )$   &    5.97   &    7.46  &    1.83  &     1.76 &   11.84 &    31.18 &    yes  \\
 G45.473$+$0.133$^{ }$    &    60.28  &  (45,55)          &  $(65   ,75  )$   &    7.61   &    5.44  &    3.51  &     3.18 &   19.35 &    23.70 &    yes  \\
 G69.540$-$0.976$^{*}$    &    11.17  &  (4,9)            &  $(13.5 ,18  )$   &    0.77   &    0.64  &    0.69  &     0.77 &    2.91 &     4.07 &    yes  \\
 G78.122$+$3.633$^{ }$    &  $-$4.00  &  ($-$18,$-$7)     &  $(1    ,11  )$   &    0.23   &    0.28  &    0.31  &     0.41 &   12.33 &    23.33 &    no   \\
 G81.722$+$0.571$^{ }$    &  $-$2.64  &  ($-$16,$-$7.5)   &  $(1.5  ,5   )$   &    0.24   &    0.24  &    0.41  &     0.52 &   25.68 &    18.64 &    yes  \\
 G81.752$+$0.591$^{ }$    &  $-$3.31  &  ($-$17,$-$8)     &  $(1    ,5   )$   &    0.52   &    0.57  &    0.52  &     0.53 &   23.49 &    19.67 &    yes  \\
 G85.410$+$0.003$^{ }$    & $-$36.50  &  ($-$49,$-$41)    &  $(-33  ,-24 )$   &    3.32   &    3.98  &    1.40  &     1.45 &    6.77 &     8.85 &    yes  \\
 G94.603$-$1.796$^{ }$    & $-$44.67  &  ($-$55,$-$47)    &  $(-41  ,-33 )$   &    1.62   &    1.35  &    0.82  &     0.50 &   10.57 &     7.16 &    no   \\
G108.184$+$5.519$^{ }$    &  $-$9.83  &  ($-$15,$-$11.5)  &  $(-7.5 ,-6  )$   &    0.08   &    0.09  &    0.20  &     0.35 &    4.03 &     2.43 &    no   \\
G109.871$+$2.114$^{ }$    &  $-$9.83  &  ($-$25,$-$17)    &  $(-5   ,7   )$   &    0.05   &    0.06  &    0.22  &     0.15 &   16.70 &    21.11 &    no   \\
G111.542$+$0.777$^{ }$    & $-$56.67  &  ($-$73,$-$64)    &  $(-47  ,-38 )$   &    0.59   &    0.74  &    0.45  &     0.61 &   12.61 &    13.29 &    no
\enddata
\tablecomments{Column (1): Outflow names, among which those with asteroids (*) are diagnosed with \coii\ line wings, else with \coi\ line wings.
Column (2): Peak velocities of either \coiii\ (signal-to-noise greater than 2) or \coii\ (if no sensitive \coiii\ detection) lines corresponding to maser velocities.
Column (3) \& (4): Velocity ranges of blue and red line wings.
Column (5) \& (6): Area of blue and red lobes.
Column (7) \& (8): Scale sizes of the major axis of blue and red lobes.
Column (9) \& (10): Integrated intensities of blue and red line wings.
Column (11): Notes for new outflows that have not been identified in the literatures. }
\end{deluxetable}

\begin{deluxetable}{lcccccccccccccr}
\tabletypesize{\scriptsize}
\tablecolumns{13}
\centering
\tabcolsep = 3.5 pt
\rotate
\tablecaption{Derived Outflow Properties\label{tab5}}
\tablewidth{0pt}
\tablehead{
\colhead{Source Name} & \colhead{$N_{\mathrm{b}}$} & \colhead{$N_{\mathrm{r}}$} &
 \colhead{$M_{\mathrm{b}}$} & \colhead{$M_{\mathrm{r}}$} & \colhead{$M_{\mathrm{out}}$} & \colhead{$p_{\mathrm{out}}$} & \colhead{$E_{\mathrm{out}}$} &
  \colhead{$t$} & \colhead{$\dot{M}_{\mathrm{out}}$} &  \colhead{$F_{\mathrm{mech}}$} & \colhead{$L_{\mathrm{mech}}$}  & \colhead{$M_{\mathrm{core}}$} \\
\colhead{} & \multicolumn{2}{c}{(10$^{20}$ cm$^{-2}$)} & \colhead{(\Msun)} & \colhead{(\Msun)} & \colhead{(\Msun)}  &  \colhead{(\Msun~\kms)} & \colhead{(10$^{45}$ erg)} & \colhead{(10$^5$ yr)} & \colhead{(10$^{-5}$ \Msun~yr$^{-1}$)}  & \colhead{(10$^{-4}$ \Msun~\kms yr$^{-1}$)} & \colhead{(\Lsun)} & \colhead{(\Msun)} \\
\colhead{(1)} & \colhead{(2)} & \colhead{(3)} & \colhead{(4)} & \colhead{(5)} & \colhead{(6)} & \colhead{(7)} & \colhead{(8)} & \colhead{(9)} & \colhead{(10)} & \colhead{(11)} & \colhead{(12)} & \colhead{(13)}
}
\startdata
%10$^{-4}$ \Msun~\kms yr$^{-1}$
 G12.199$-$0.033$^{ }$  & \nodata & 1.4     & \nodata & 26       & 26   & 100  & 4.7     & 42  &  6.2   &  2.5    &  0.09     & 6500   \\
 G12.681$-$0.182$^{*}$  & \nodata & 13.7    & \nodata & 13       & 13   & 40   & 1.4     & 13  &  9.8   &  3      &  0.08     & 1500   \\
 G12.889$+$0.489$^{ }$  & \nodata & 1.8     & \nodata & 1.5      & 1.5  & 10   & 0.8     & 11  &  1.3   &  0.92   &  0.06     & 1100   \\
 G14.101$+$0.087$^{ }$  & 1.4     & \nodata & 9.4     & \nodata  & 9.4  & 49   & 2.7     & 30  &  3.1   &  1.7    &  0.08     & 1400   \\
 G14.332$-$0.639$^{ }$  & 1.8     & \nodata & 0.23    & \nodata  & 0.23 & 1.8  & 0.2     & 1.8 &  1.2   &  1      &  0.07     & 110    \\
 G16.585$-$0.051$^{*}$  & 22.0    & 15.8    & 57      & 34       & 91   & 230  & 6.5     & 35  &  26    &  6.7    &  0.2      & 1800   \\
 G16.864$-$2.159$^{ }$  & 2.2     & 1.9     & 1.9     & 1        & 2.9  & 17   & 1.1     & 8.3 &  3.5   &  2      &  0.1      & 82     \\
 G17.638$+$0.157$^{*}$  & 20.3    & 19.2    & 21      & 20       & 41   & 210  & 12.1    & 17  &  25    &  13     &  0.6      & 290    \\
 G18.461$-$0.004$^{*}$  & 26.5    & 41.5    & 60      & 260      & 320  & 1300 & 58.8    & 38  &  86    &  34     &  1        & 1500   \\
 G19.365$-$0.030$^{*}$  & 26.5    & 10.1    & 25      & 9.4      & 34   & 100  & 3.5     & 24  &  14    &  4.4    &  0.1      & 550    \\
 G20.081$-$0.135$^{*}$  & 10.0    & 12.0    & 51      & 120      & 170  & 690  & 30.5    & 54  &  32    &  13     &  0.5      & 3100   \\
 G23.010$-$0.411$^{ }$  & \nodata & 2.8     & \nodata & 20       & 20   & 180  & 19.3    & 13  &  15    &  14     &  1        & 8000   \\
 G23.440$-$0.182$^{ }$  & \nodata & 1.6     & \nodata & 7.6      & 7.6  & 110  & 15.9    & 6   &  13    &  18     &  2        & 8200   \\
 G23.484$+$0.097$^{ }$  & 1.8     & \nodata & 5.5     & \nodata  & 5.5  & 45   & 4.0     & 9.3 &  5.9   &  4.9    &  0.4      & 1800   \\
 G24.493$-$0.039$^{*}$  & 36.6    & 18.8    & 350     & 160      & 510  & 2200 & 100.0   & 28  &  180   &  77     &  3        & 3400   \\
 G24.790$+$0.083$^{*}$  & \nodata & 22.8    & \nodata & 540      & 540  & 2000 & 83.7    & 83  &  65    &  24     &  0.8      & 23000  \\
 G25.411$+$0.105$^{*}$  & 6.8     & 18.0    & 18      & 78       & 95   & 360  & 16.0    & 36  &  27    &  10     &  0.4      & 570    \\
 G25.650$+$1.050$^{ }$  & 2.1     & 1.7     & 66      & 55       & 120  & 1700 & 246.0   & 27  &  45    &  62     &  7        & 63000  \\
 G25.826$-$0.178$^{ }$  & 1.6     & \nodata & 6.8     & \nodata  & 6.8  & 69   & 7.6     & 12  &  5.5   &  5.6    &  0.5      & 710    \\
 G26.602$-$0.220$^{ }$  & 0.6     & 0.6     & 7.1     & 7.4      & 14   & 44   & 1.5     & 49  &  3     &  0.9    &  0.03     & 990    \\
 G27.223$+$0.137$^{ }$  & 0.4     & 0.4     & 4.4     & 2.2      & 6.7  & 21   & 0.7     & 40  &  1.7   &  0.51   &  0.02     & 290   \\
 G27.286$+$0.151$^{ }$  & 1.7     & \nodata & 45      & \nodata  & 45   & 270  & 16.6    & 60  &  7.5   &  4.5    &  0.2      & 8800   \\
 G27.369$-$0.164$^{ }$  & 1.0     & 1.8     & 11      & 27       & 37   & 160  & 7.4     & 41  &  9.1   &  3.8    &  0.2      & 6900   \\
 G28.201$-$0.049$^{*}$  & 16.4    & \nodata & 340     & \nodata  & 340  & 1000 & 32.8    & 72  &  48    &  14     &  0.4      & 5200   \\
 G28.829$+$0.488$^{*}$  & 13.6    & 7.8     & 300     & 220      & 530  & 2700 & 163.0   & 85  &  62    &  32     &  2        & 1200   \\
 G28.832$-$0.253$^{*}$  & 15.3    & \nodata & 81      & \nodata  & 81   & 280  & 10.4    & 33  &  25    &  8.5    &  0.3      & 2400   \\
 G29.863$-$0.044$^{ }$  & \nodata & 5.3     & \nodata & 14       & 14   & 100  & 9.0     & 27  &  5     &  3.8    &  0.3      & 7800   \\
 G31.282$+$0.062$^{ }$  & \nodata & 0.8     & \nodata & 2.9      & 2.9  & 19   & 1.3     & 21  &  1.4   &  0.89   &  0.05     & 1500   \\
 G35.197$-$0.743$^{*}$  & 23.0    & 11.1    & 52      & 27       & 79   & 380  & 21.4    & 19  &  42    &  20     &  0.9      & 1400   \\
 G40.623$-$0.138$^{*}$  & 8.3     & 10.5    & 7.1     & 9        & 16   & 60   & 2.4     & 13  &  12    &  4.6    &  0.2      & 87     \\
 G41.348$-$0.136$^{ }$  & 0.6     & 0.6     & 7.8     & 11       & 18   & 62   & 2.2     & 110 &  1.7   &  0.58   &  0.02     & 700    \\
 G43.166$-$0.002$^{ }$  & 3.0     & \nodata & 110     & \nodata  & 110  & 970  & 94.7    & 45  &  25    &  22     &  2        & 3700   \\
 G43.796$-$0.127$^{*}$  & 9.0     & 10.3    & 65      & 75       & 140  & 570  & 25.7    & 57  &  25    &  10     &  0.4      & 640    \\
 G43.890$-$0.784$^{ }$  & 2.9     & 2.9     & 40      & 40       & 80   & 610  & 53.9    & 26  &  31    &  24     &  2        & 9400   \\
 G45.467$+$0.053$^{ }$  & 1.8     & 4.7     & 17      & 56       & 73   & 540  & 49.7    & 19  &  38    &  28     &  2        & 5000   \\
 G45.473$+$0.133$^{ }$  & 2.9     & 3.6     & 35      & 31       & 66   & 470  & 34.6    & 48  &  14    &  9.7    &  0.6      & 2500   \\
 G69.540$-$0.976$^{*}$  & 8.5     & 11.9    & 10      & 12       & 22   & 67   & 2.1     & 25  &  8.9   &  2.6    &  0.07     & 520    \\
 G78.122$+$3.633$^{ }$  & 1.9     & 3.5     & 0.67    & 1.6      & 2.3  & 18   & 1.7     & 5.3 &  4.3   &  3.5    &  0.3      & 130    \\
 G81.722$+$0.571$^{ }$  & 3.9     & 2.8     & 1.5     & 1.1      & 2.5  & 16   & 1.1     & 8.2 &  3.1   &  1.9    &  0.1      & 460    \\
 G81.752$+$0.591$^{ }$  & 3.6     & 3.0     & 3       & 2.7      & 5.7  & 36   & 2.5     & 8.2 &  6.9   &  4.4    &  0.3      & 230    \\
 G85.410$+$0.003$^{ }$  & 1.0     & 1.3     & 5.4     & 8.5      & 14   & 81   & 5.1     & 24  &  5.8   &  3.4    &  0.2      & 410    \\
 G94.603$-$1.796$^{ }$  & 1.6     & 1.1     & 4.1     & 2.3      & 6.4  & 30   & 1.6     & 16  &  4     &  1.9    &  0.08     & 140    \\
G108.184$+$5.519$^{ }$  & 0.6     & 0.4     & 0.07    & 0.05     & 0.13 & 0.3  & 0.008   & 14  &  0.09  &  0.022  &  0.0004   & 9      \\
G109.871$+$2.114$^{ }$  & 2.5     & 3.2     & 0.21    & 0.32     & 0.52 & 4.6  & 0.5     & 2.3 &  2.2   &  2      &  0.2      & 30     \\
G111.542$+$0.777$^{ }$  & 1.9     & 2.0     & 1.8     & 2.4      & 4.2  & 48   & 5.7     & 5.3 &  7.9   &  9      &  0.9      & 1300
\enddata
\tablecomments{Derived outflow parameters without inclination correction.
Column (1): Outflow names with signs (*) for \coi\ or \coii\ outflows, same as Table~\ref{tab4}.
Column (2) \& (3): \Hm\ gas column densities of blue and red lobes derived from either Equation~(\ref{equ_N12}) or (\ref{equ_N13}) multiplying a conversion factor.
Column (4) -- (6): The blue/red lobes and total outflow masses of \Hm\ gas derived from Equation~(\ref{equ_mbr}).
Column (7) -- (12): Outflow momenta, energies, dynamical time scales, mass rates, mechanical forces and luminosities, derived from Equation~(\ref{equ_p}), (\ref{equ_E}), (\ref{equ_t}), (\ref{equ_m/t}), (\ref{equ_Fm}) and (\ref{equ_Lm}), respectively.
Column (13): \coiii\ core masses derived from Equation~(\ref{equ_N18}) and (\ref{equ_M18}).
}
\end{deluxetable}

\end{document}